\documentclass[%
 twocolumn, amsmath,amssymb,
 aps,prd, showpacs
]{revtex4-2}

\usepackage{graphicx}
\usepackage{dcolumn}
\usepackage{bm}
\usepackage{lineno}
\usepackage{amsmath}
\usepackage{relsize}
\usepackage{color}
\usepackage{here}

\begin{document}


\title{Testing unified models for the origin of ultrahigh-energy cosmic rays and neutrinos: \\ Multimessenger approaches with x-ray observations}

\author{Shigeru Yoshida}
 \email{syoshida@hepburn.s.chiba-u.ac.jp}
\affiliation{%
International Center for Hadron Astrophysics, Chiba University, Chiba 263-8522 Japan}


\author{Kohta Murase}
\thanks{murase@psu.edu}
\affiliation{Department of Physics; Department of Astronomy \& Astrophysics; Center for Multimessenger Astrophysics, Institute for Gravitation and the Cosmos, 
The Pennsylvania State University, University Park, Pennsylvania 16802, USA\\
and \\
Center for Gravitational Physics and Quantum Information, Yukawa Institute for Theoretical Physics, Kyoto University, Kyoto, Kyoto 606-8502, Japan}

\date{\today}

\begin{abstract}
The unified models of astrophysical sources to account for ultrahigh-energy cosmic rays (UHECRs) and high-energy cosmic neutrinos with energies greater than 100~TeV have been discussed. 
Based on model-independent arguments, we argue that if the photomeson production is the dominant mechanism, the most probable candidate sources are x-ray transient objects, allowing for the semitransparency for the photomeson production. 
We develop a generic model of high-energy neutrino emitters accompanied by x-ray emission, and present how multimessenger observations can place significant constraints on the source parameters that characterize the common sources of neutrinos and UHECRs,  
such as the cosmic-ray loading factor. 
The requirements of UHECR acceleration, escape, and energetics further constrain the magnetic field and the bulk Lorentz factor of the sources. 
The resulting bounds provide diagnoses of the unified models, which demonstrates the importance of current and future x-ray facilities such as {\it MAXI} and {\it Einstein Probe}.
\end{abstract}


\maketitle


\section{\label{sec:introduction} Introduction}

The cosmic background radiation in high-energy and ultrahigh-energy (UHE) sky at $\gtrsim 1$~PeV ($10^{15}$~eV) is formed by cosmic rays and neutrinos. The precise measurements of
ultrahigh-energy cosmic rays (UHECRs) by the Pierre Auger Observatory (PAO)
with high statistics have now revealed the detailed structure of their energy spectrum~\cite{Aab:2016zth}.
The IceCube Neutrino Observatory has discovered~\cite{Aartsen:2013bka, Aartsen:2013jdh} and measured the high-energy neutrino radiation in the UHE sky~\cite{Aartsen:2020aqd, IceCube:2021uhz}, realizing the observation window of the penetrating messengers to study the UHE emissions. It has been widely recognized that neutrino detection is a smoking gun for identifying the origins of high-energy cosmic rays, as high-energy neutrinos can mainly be produced by the hadronic processes involving high-energy nucleons or nuclei, while high-energy electromagnetic (EM) emissions can naturally be realized by the leptonic processes originating in electrons or positrons and magnetic fields. The reality is not that simple, however. The existence of high-energy protons or nuclei does not necessarily imply cosmic-ray emission because these charged particles must escape from their acceleration site, which is not guaranteed to be possible. Any dense environment formed by radiation or matter within or in the vicinity of the acceleration region causes significant energy losses to prevent radiation of cosmic rays. It leads to neutrino emission not accompanied by escaping cosmic rays.

The recent discovery of neutrino emission from the Seyfert galaxy NGC 1068~\cite{Aartsen:2019fau, IceCube:2022der} indeed belongs to this case, as the parent protons to produce neutrinos are unlikely to run away from the inner region close to the central supermassive black hole~\cite{Murase:2022dog}.
Moreover, the energy range of cosmic background neutrinos discovered by IceCube is mostly below 10~PeV ($10^{16}$~eV), which is orders of magnitude lower than the main extragalactic UHECR energies ($\gtrsim 10^{19}$~eV), and it is not obvious that the measured neutrinos are associated with UHECRs. 
The question of whether the two primary components of UHE sky, cosmic rays, and neutrinos, are relevant to each other has not yet been answered. There are various candidates of UHECR sources, including active galactic nuclei (AGNs), gamma-ray bursts (GRBs), and tidal disruption events (TDEs)~\cite{AlvesBatista:2019tlv}.

\begin{figure}[t]
\begin{center}
\includegraphics[width=0.45\textwidth]{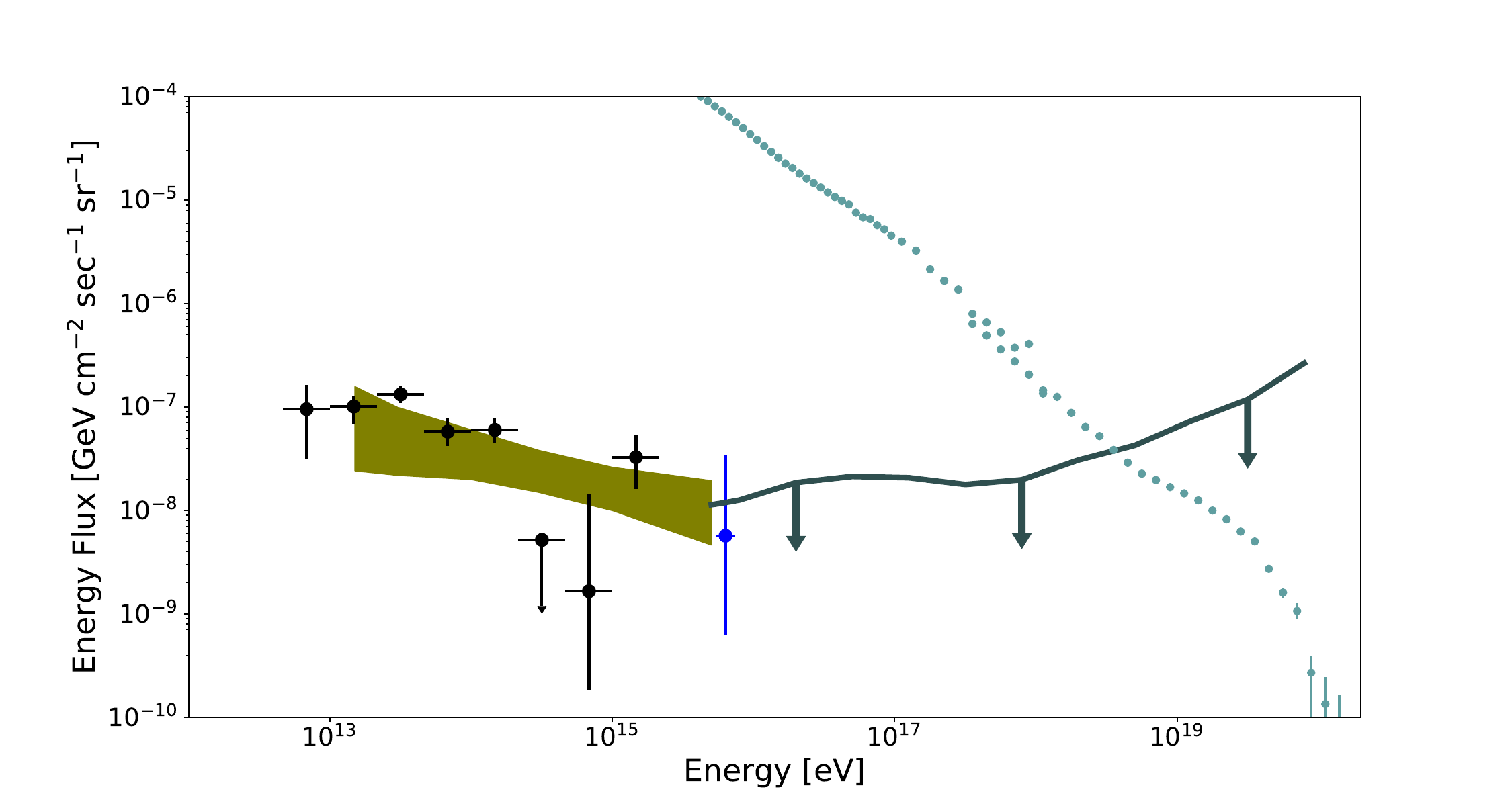}
\caption{Energy fluxes of high-energy cosmic background radiations. 
    The small data points represent the UHECR fluxes measured by PAO~\cite{Aab:2016zth} and IceTop~\cite{Aartsen:2013wda}.
    The rest of the data shows the neutrino energy fluxes and
    the upper limits measured with IceCube. 
    The thick black data points were obtained by the neutrino-induced cascade measurements~\cite{Aartsen:2020aqd}. The shaded region indicates the flux space consistent with the $\nu_\mu$-induced track measurements~\cite{IceCube:2021uhz}. 
    The blue data point shows the flux of the 6 PeV-energy $\bar{\nu_e}$ event estimated by the dedicated search channel called PEPE (PeV Energy Partially-contained Events)~\cite{IceCube:2021rpz}.
    The thick line with arrows indicates the differential upper limit obtained by
    the extremely high-energy (EHE) neutrino search~\cite{Aartsen:2018vtx}.
    The neutrino fluxes are the all-flavor-sum fluxes $E_\nu^2 \Phi_{\nu_e + \nu_\mu + \nu_\tau}$.
 }
\label{fig:energy_fluxes}
\end{center}
\end{figure}

As shown in Fig.~\ref{fig:energy_fluxes}, the observed energy flux of high-energy neutrinos above $~\sim 100$~TeV is comparable to that of UHECRs at $\gtrsim 10^{19}$~eV. This suggests that neutrino background radiation may originate from the same class of astrophysical objects to produce UHECRs. It may be plausible that the cosmic-ray radiation from UHECR accelerators can be understood in a common unified scheme and even the connection with the gamma-ray background radiation may be expected~\cite{Murase:2016gly,Murase:2018utn}.

Motivated by this observational fact, Yoshida and Murase (YM20 \cite{Yoshida:2020div}) built the generic unification model to account for the observed neutrinos at energies greater than $\sim 100$~TeV and UHECRs, based on the photohadronic interaction framework. The UHECR luminosity density and the optical depth to the photomeson production are constrained by the measured UHECR and neutrino fluxes and the other conditions
required to emit UHECR nucleons or nuclei. The generic constraints
enable us to evaluate whether a given class of astronomical objects is qualified as the common origin of UHECRs and neutrinos, 
which will be useful as a probe of the origin of UHECRs with multimessenger observations.

In this paper, we discuss which objects among the known astronomical object classes meet the criteria for UHECR accelerators in the unified UHE particle emission scheme. We deconstruct the generic constraints obtained by YM20 into a set of straightforward analytical formulas, which allows us to diagnose the likelihood of a given object being a candidate of the common sources. We argue that the most plausible candidates are x-ray transients allowing the semitransparency for the photomeson production. We then construct a generic modeling of neutrino emitters with optically thin x-ray targets by the unification scheme of neutrino and UHECR origins based on YM20. The results will be described in the parametrization to characterize the sources, such as the source number density, the Lorentz bulk factor of plasma, and the cosmic-ray loading factor, for a given source luminosity in the x-ray band. 
We present how current and future multimessenger observations involving x rays constrain these parameters, by which we pin down the unified origin or fully rule out the UHE unification scenario in the photohadronic framework. 

This paper is organized as follows. In Sec.~\ref{sec:generic_model}, we revisit the framework of YM20~\cite{Yoshida:2020div}, and introduce the analytical formulas to describe necessity conditions for common UHECR and neutrino sources. In Sec.~\ref{sec:case_study}, we evaluate each of the known astronomical object classes 
that have been considered as UHECR sources in the literature to see their applicability to the unified source scheme. 
In Sec.~\ref{sec:neutrino_xray}, we introduce a generic x ray -- neutrino -- UHECR source model with a set of parameters observable by the present or future x-ray missions. We present constraints in the parameter space placed by x-ray observations and various UHECR source conditions.
We summarize the implications of the present constraints and future perspective in Sec.~\ref{sec:discussion}.
The standard cosmology with $H_0 \simeq 73.5$ km $\sec^{-1}$ Mpc$^{-1}$,
$\Omega_M = 0.3$, and $\Omega_{\Lambda}=0.7$ is assumed throughout the paper. Primed (') characters represent quantities in the comoving frame of plasma outflow.

\section{\label{sec:generic_model} Generic unified model}

\begin{figure*}[t]
\begin{center}
\includegraphics[width=0.4\textwidth]{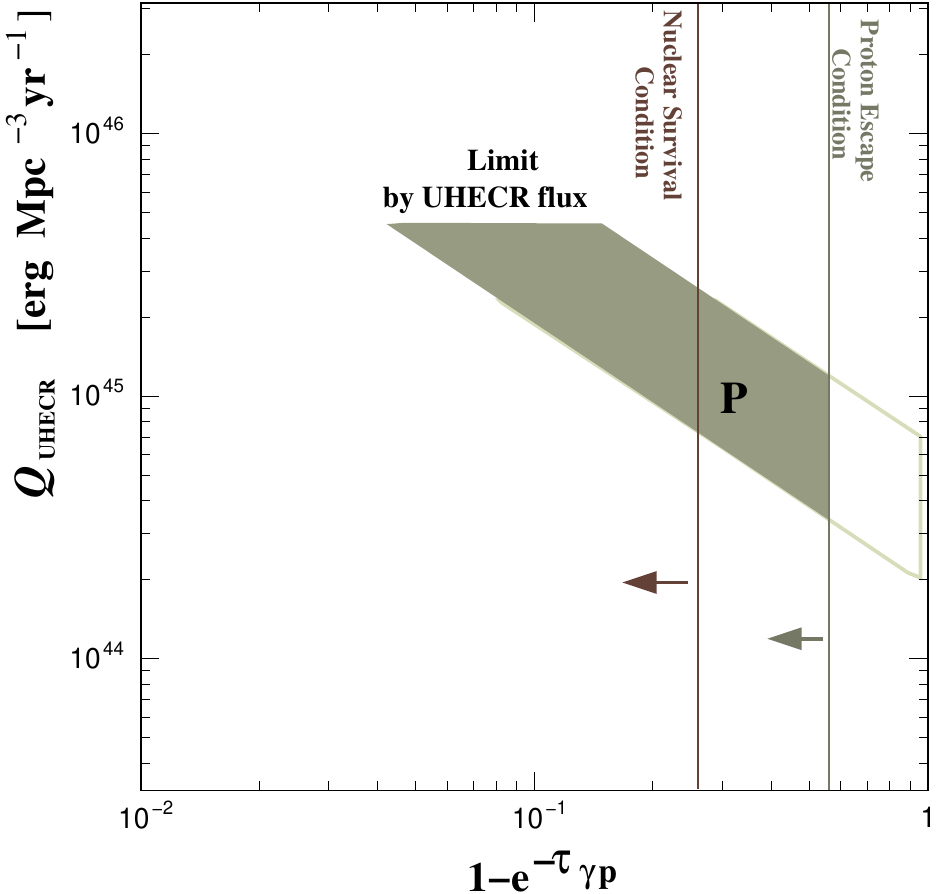}
\includegraphics[width=0.4\textwidth]{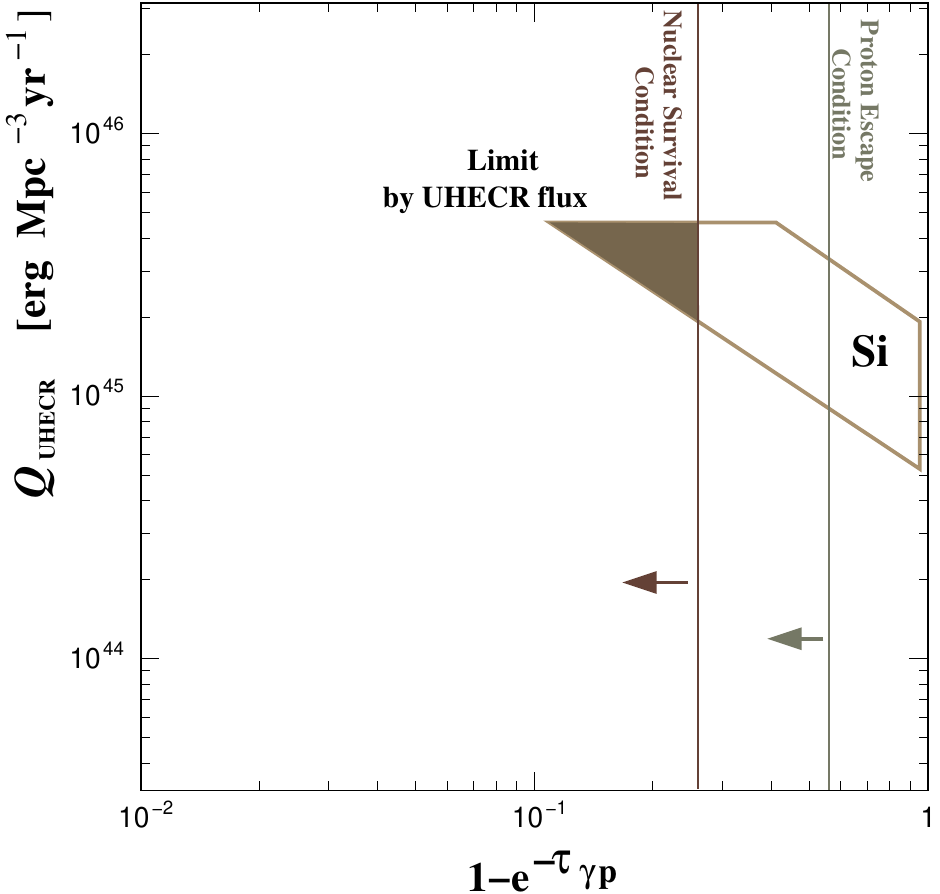}
\caption{The allowed region in the parameter space of luminosity density, $Q_{\rm UHECR}$, and damping factor $\displaystyle{1-e^{-\tau_{p\gamma0}}}$~\cite{Yoshida:2020div}.
The region enclosed by the solid lines displays the allowed space by the UHECR and the neutrino flux requirements. The shaded region represents the parameter space allowed also by considering the UHECR proton escape condition or the nuclear survival condition. 
The left panel shows the proton model, while the right panel shows the case of primary silicon nuclei.}
\label{fig:unification_scenario_constraints}
\end{center}
\end{figure*}

In YM20, a nonthermal photon field represented by a power-law spectrum filling the acceleration and/or the emission region in a source
provides a target for UHECRs to yield secondary neutrinos
via photohadronic ($p\gamma$) interactions. The $p\gamma$ interaction
efficiency is parametrized by the $p\gamma$ optical depth, $\tau_{p\gamma0}$, at the fiducial UHECR energy $\varepsilon_{\rm UHECR}^{\rm FID}=10$~PeV.
The $p\gamma$ optical depth is generically given by
\begin{eqnarray}
\tau_{p\gamma}(\varepsilon_p) &\approx & \tau_{p\gamma0} {\left(\frac{\varepsilon_p}{{\varepsilon}_{\rm UHECR}^{\rm FID}}\right)}^{\alpha_\gamma-1} \nonumber \\
&\approx & \frac{B'}{\Gamma^2}\sqrt{\frac{L'_{\gamma}}{\xi_{\rm B}}} \beta^{-1} C(\alpha_\gamma)
{\left(\frac{\varepsilon_p}{{\varepsilon}_{\rm UHECR}^{\rm FID}}\right)}^{\alpha_\gamma-1}.
\label{eq:optical_depth_general}
\end{eqnarray}
Here $\Gamma$ is the bulk Lorentz factor of the plasma in the interaction site, $\beta$ is the plasma velocity, $\alpha_\gamma$ is the power-law index of the target photon spectrum, $B'$ is the magnetic field strength measured in the plasma rest frame, and $L'_\gamma$ is the target photon luminosity in the plasma rest frame.
The magnetic field loading factor, $\xi_{\rm B}$, is the ratio of the magnetic energy density $U^{'}_{\rm B}$ to the photon radiation energy density and defined by
\begin{eqnarray}
\xi_{\rm B} &\equiv& U'_{\rm B}\left(\frac{L_\gamma}{4\pi \Gamma^2 R^2c}\right)^{-1}.
\label{eq:magnetic_energy_density}
\end{eqnarray}
The coefficient $C(\alpha_\gamma)$ in Eq.~(\ref{eq:optical_depth_general})
is determined by the photomeson production cross section and the target photon spectral shape, and given by
\begin{eqnarray}
C(\alpha_\gamma)&\sim&2.4\times{10}^{-24}~{\rm erg}^{-1}~{\rm cm}^{3/2}~{\rm s}^{1/2}\nonumber\\
&\times&\left(\frac{2}{1+\alpha_\gamma}\right).
\end{eqnarray}
The necessity conditions to be qualified for UHECR emitters
can also be formulated using the optical depth, which we describe later.

For a given $\tau_{p\gamma0}$, the analytical formulation
to calculate the neutrino flux has been derived in YM20. Using the analytical representation of the UHECR flux taking into account
the energy loss effect during the propagation in the cosmic microwave background (CMB) field filling the intergalactic space, the model has constrained the parameter space of $\tau_{p\gamma0}$ and the integrated luminosity density of UHECRs in the local universe $Q_{\rm UHECR}$ by comparisons to the measured UHECR and neutrino fluxes. Figure~\ref{fig:unification_scenario_constraints}
displays the resultant constraints obtained by YM20, for the two cases when the extragalactic UHECRs are dominated by protons, and dominated by nuclei, considering silicon as a benchmark, respectively.

Our goal here is to interpret these constraints 
by a set of simple approximated formulas
to quickly investigate a given astronomical object for
qualifications as the UHECR-neutrino common source.
We start with the UHECR energetics argument.
The UHECR differential luminosity density is estimated as (e.g., Ref.~\cite{Murase:2018utn} and references therein)
\begin{eqnarray}
    E_{\rm CR}\frac{dQ_{\rm UHECR}}{dE_{\rm CR}} &\approx& 6.3\times 10^{43}~{\rm erg\ Mpc^{-3} yr^{-1}}\nonumber\\
    &\times&
    \left(\frac{E_{\rm CR}}{10^{19.5}\ {\rm eV}}\right)^{2-\alpha_{\rm CR}}
\end{eqnarray}
where $\alpha_{\rm CR}$ is the power-law index of UHECR spectrum.
The integrated luminosity density of UHECRs above the energy of 
$\varepsilon_p\geq \varepsilon_{\rm UHECR}^{\rm FID} (=10~{\rm PeV)}$ is obtained by the transformation from the differential luminosity density above and given by
\begin{eqnarray}
    Q_{\rm UHECR}
    &\approx& 2.3\times 10^{45}~{\rm erg \ Mpc^{-3} yr^{-1}}\nonumber\\
    &\approx& n_0^{\rm eff}\xi_{\rm UHECR}L_\gamma
    \label{eq:UHECR_energetics}
\end{eqnarray}
when we use $\alpha_{\rm CR}=2.3$ as a benchmark.
Here, the UHECR loading factor, $\xi_{\rm UHECR}$, is the ratio 
of the UHECR bolometric luminosity $L_{\rm UHECR}$ and 
the target photon luminosity $L_\gamma$,
and $n_0^{\rm eff}$ is the effective local source number density at redshift $z=0$. 
Note that this effective number density can be different from the number density measured by UHECR observations. This is because cosmic rays are intervened by magnetic fields and the number density in the cosmic-ray sector should be affected by time delays~\cite{Miralda-Escude:1996twc,Murase:2008sa}.
One can see that this is consistent with the range of y axis
in the allowed region shown in Fig.~\ref{fig:unification_scenario_constraints}.
This energetics condition above effectively sets the requirement of the UHECR loading factor for a given $L_\gamma$ and $n_0^{\rm eff}$, such as
\begin{equation}
    \xi_{\rm UHECR} \sim 0.7 \left(\frac{L_\gamma}{10^{46} {\rm erg/s}}\right)^{-1}\left(\frac{n_0^{\rm eff}}{10^{-8} {\rm Mpc^{-3}}}\right)^{-1}.
    \label{eq:cr_loading}
\end{equation}
Note that $L_\gamma$ should be evaluated at the representative energy band, where UHECRs with $\varepsilon_{\rm UHECR}^{\rm FID}(=10~{\rm PeV})$ 
mainly interact via the $\Delta(1232)$ resonance of photohadronic collisions [see Eq.~(12) in YM20 and Eq.~(\ref{eq:photon_energy_ref}) in Sec.~\ref{subsec:generic_model_xray}], 
and $n_0^{\rm eff}$ is determined at the local density represented by
this luminosity [see Eq.~9 of Ref.~\cite{Murase:2016gly}]. 
In the standard candle source approximation, it is identical to the net local source density.

Next, we derive the requirements by the all-sky neutrino flux.
The neutrino all-flavor-sum emissivity from a source is connected to the primary UHECR emissivity as~\cite{Murase:2015xka}
\begin{equation}
  \varepsilon_\nu^2 \frac{d\dot{N}_{\nu}}{d\varepsilon_\nu} \approx \xi_\pi \langle y_\nu \rangle \langle x \rangle  \tau_{p\gamma}\varepsilon_{\rm CR}^2\frac{d\dot{N}_{\rm CR}}{d\varepsilon_{\rm CR}} 
  \label{eq:nuToCr}
\end{equation}
for a hadronically thin ({\it i.e.}$\tau_{p\gamma}\lesssim 5$) source. 
Here $\langle x\rangle\sim 0.2$ is the average inelasticity of the $p\gamma$ collision, and $\langle x\rangle \tau_{p\gamma}$ corresponds to the effective optical depth ($f_{p\gamma}$) used in the literature.
Also, $\langle y_\nu\rangle\sim 1/4$ is the average fraction
of energy channeling into a neutrino from the secondary produced pion, and
$\xi_\pi\sim 3/2$ is the average multiplicity of neutrinos from a single pion produced by the photohadronic interaction considering the $\pi^+:\pi^0$ ratio is $\approx1:1$.
Since the energy flux of the high-energy cosmic background neutrinos can be approximately written using the source emissivity, $\displaystyle \varepsilon_\nu^2 d\dot{N}_{\nu}/d\varepsilon_\nu$,
we can relate the $p\gamma$ optical depth to the required bolometric UHECR luminosity density for a given energy flux of neutrinos via Eq.~(\ref{eq:nuToCr}). We get
\begin{eqnarray}
\tau_{p\gamma0}Q_{\rm UHECR} &\approx& 9.3\times 10^{43} \left(\frac{E_\nu^2\Phi_\nu}{2\times 10^{-8}\ [{\rm GeV\ cm^{-2} s^{-1} sr^{-1}]}}\right)\nonumber \\
 && \times\left(\frac{\xi_z}{2.8}\right)^{-1}
 \quad {\rm erg\ Mpc^{-3} yr^{-1}},
 \label{eq:neutrino_flux}
\end{eqnarray}
for $\alpha_{\rm CR}=2.3$. Here $\displaystyle \xi_z\equiv (1/t_{\rm H})\int dt \Psi(z)/(1+z)$ is
a dimensionless parameter that depends on the redshift evolution function $\Psi(z)$ of the sources.
This relation well represents the allowed parameter space shown in Fig.~\ref{fig:unification_scenario_constraints}.

Combining the UHECR luminosity density given by Eq.~(\ref{eq:UHECR_energetics}) with
this neutrino fiducial flux condition, Eq.~(\ref{eq:neutrino_flux}), sets the lower bound
of the $p\gamma$ optical depth, which is~\cite{Yoshida:2014uka}
\begin{eqnarray}
    \tau_{p\gamma0}&\gtrsim& 0.04 
    \left(\frac{\xi_z}{2.8}\right)^{-1}. 
    \label{eq:neutriono_flux_requirement}
\end{eqnarray}

Third, we consider the conditions for the acceleration of UHECRs.
The power required from the Hillas condition is represented as the condition of $L'_\gamma$ as a function of the maximal accelerated cosmic-ray energy $\varepsilon_i^{\rm max}$ (e.g., Refs.~\cite{Lemoine:2009pw,Yoshida:2020div,AlvesBatista:2019tlv}).
This condition can be translated into the lower bound of the magnetic field loading factor $\xi_B$ defined as Eq.~(\ref{eq:magnetic_energy_density}).
We get
\begin{eqnarray}
\xi_B &\geq& \frac{1}{2} c\eta^2\beta^2L^{'-1}_\gamma\left(\frac{\varepsilon_i^{\rm max}}{Z e}\right)^2  \nonumber\\
&\gtrsim&1.7\eta^{2}\beta^2\left(\frac{L_\gamma}{10^{46} {\rm erg/s}}\right)^{-1}
\left(\frac{\Gamma}{10^{0.5}}\right)^2
\left(\frac{\varepsilon_i^{\rm max}}{Z10^{11}~{\rm GeV}}\right)^2,\nonumber\\
\label{eq:hillas_condition}
\end{eqnarray}
where $\eta\gtrsim 1$ represents the efficiency of the particle acceleration,
and the limit of $\eta\rightarrow \beta^{-2}$ corresponds to the Hillas condition in the case of diffusive shock acceleration~\cite{AlvesBatista:2019tlv}. In addition, to ensure that UHECRs can leave the sources before losing their energies, the escape timescale must be faster than the cosmic-ray energy loss timescale.
The energy loss processes consist of the $p\gamma$ photomeson production,
Bethe-Heitler (BH) interactions, and the synchrotron cooling.
The photomeson production timescale is essentially counted with the $p\gamma$ optical depth, $\tau_{p\gamma}$, in the present scheme, and any sources with $\tau_{p\gamma}(\varepsilon_i^{\rm max})\lesssim 5$
implies that the energy losses by the photomeson production
is not a deciding factor to limit the UHECR acceleration and escape processes. As the BH process is in general important only if the photon spectrum is softer as $\alpha_\gamma\gtrsim 2$, the UHECR escape condition is formulated as a necessity condition by requiring
the dynamical timescale faster than the synchrotron cooling timescale.
It has been found that this condition is transformed into the upper bound
$\tau_{p\gamma0}$ as~\cite{Yoshida:2020div}
\begin{eqnarray}
\tau_{p\gamma0} &\lesssim & 0.06 \frac{2}{1+\alpha_\gamma} \xi_B^{-1}{\beta^{-1}\left(\frac{A}{Z}\right)}^4 {\left(\frac{\varepsilon_i^{\rm max}}{10^{11}\ {\rm GeV}}\right)}^{-1}.
\label{eq:esc_condition}
\end{eqnarray}
Note that the diffusion motion of cosmic rays in the photohadronic interaction zone makes this condition even more stringent. Thus, the bound placed by Eq.~(\ref{eq:esc_condition}) above can be regarded as conservative.

If the bulk of the measured UHECRs is dominated by heavier nuclei rather than nucleons as strongly indicated by the PAO data, 
the further severe requirement must be satisfied -- the nuclei survival condition~\cite{Murase:2010gj}.
We require that nuclei with $A > 1$ and $Z > 1$ are accelerated and survive. This is possible only if the timescale of the photodisintegration is slower than the dynamical timescale, leading to the condition on photodisintegration optical depth as $\tau_{A\gamma} \lesssim A$. We find that this condition sets the upper bound
of the $p\gamma$ optical depth as~\cite{Yoshida:2020div}
\begin{eqnarray}
\tau_{p\gamma0} &\lesssim& A\frac{\int ds \frac{\sigma_{p\gamma}(s)}{s-m_p^2}}
{\int ds \frac{\sigma_{A\gamma}(s)}{s-m_A^2}}
{\left[\left(\frac{s_{\rm GDR}-m_A^2}{s_\Delta-m_p^2}\right)
\left(\frac{\varepsilon_p^{\rm 10 PeV}}{\varepsilon_{i}^{\rm max}}\right)\right]}^{\alpha_\gamma-1} \nonumber \\
&\lesssim & 0.4~{\left(\frac{A}{56}\right)}^{-0.21},
\label{eq:suvival_condition}
\end{eqnarray}
where $s_\Delta$ is the Mandelstam variable at the $\Delta$ resonance 
of the photomeson interactions and $s_{\rm GDR}$ is the Mandelstam variable at the giant dipole resonance of the photodisintegration.
This is equivalent to the effective nucleus-survival bound obtained by Ref.~\cite{Murase:2010gj}.

\section{\label{sec:case_study} Diagnosis of candidate sources for the unified model}

The series of equations to describe the conditions for accounting 
for the UHECRs and $0.1-1$~PeV energy neutrino measurements discussed in the previous section are generic without relying on specific model-dependent arguments, in exchange for more conservative constraints compared to those imposed by a model dedicated for a given object. In other words, a class of sources failing to meet the criteria 
cannot be the unified source without introducing more complicated modeling such as multizones for cosmic ray accelerations and neutrino productions. Here we run the generic diagnosis to check if any known class of astronomical object remains in a possible candidate to unify the picture of UHE astroparticle radiation.

\subsection{Steady sources}
%
\begin{table*}
    \caption{Parameters characterizing UHECR/neutrino/EM emission
    and the resultant constraints on the UHECR loading factor, $\xi_{\rm UHECR}$, the magnetic field parameter $\xi_{\rm B}$, and the photohadronic optical depth at 10~PeV $\tau_{p\gamma0}$, imposed by the conditions for UHECR acceleration, the fiducial neutrino flux, UHECR escape, and UHECR nucleus survival.
    The various sites/populations in the AGN family are listed.
    Note that parameters have significant uncertainties although only specific parameters are shown in this table. We use $\varepsilon_i^{\rm max}=10^{11}$~GeV. 
    \label{table:constraints_AGN}}
\begin{ruledtabular}
  \begin{tabular}{lcccc}
    &  RL AGN (BL Lac jet) & RL AGN (FSRQ jet) & RQ AGN (jet) & RL AGN (hot disk)\\
    \hline
    $\Gamma\beta$ of the outflow 
    & $\sim$~10 
    & $\sim$~10 & 
    $\sim1$ 
    & $\sim0.01$ \\
    Target photon energy 
    & UV/x-ray 
    & opt/UV 
    & opt/UV
    & IR/opt\\
    $L_\gamma^{\rm eff}\ [{\rm erg/s}]$ 
    & ${\rm a~few}\times 10^{45}$ 
    & ${\rm a~few}\times 10^{47}$ 
    & ${\rm a~few}\times 10^{43}$ 
    & ${\rm a~few}\times10^{41}$ \\
    
    $n_0^{\rm eff} \ [{\rm Mpc^{-3}}]$ 
    & $\sim10^{-9}$ & $\sim10^{-11}$ & $\sim10^{-6}$ & $\sim10^{-7}$\\
    $n_0^{\rm tot} \ [{\rm Mpc^{-3}}]$ &  $\sim10^{-7}-10^{-6}$ 
    & $\sim10^{-9}-10^{-8}$ 
    & $\sim10^{-4}-{10}^{-3}$ 
    & $\sim10^{-5}-{10}^{-4}$\\
    \hline
    
    $R\  [{\rm cm}]$ 
    & ${\rm a~few}\times{10}^{17}$ 
    & ${\rm a~few}\times{10}^{17}$ 
    & ${\rm a~few}\times{10}^{18}$ 
    & ${\rm a~few}\times{10}^{14}$\\
    $B'\ [{\rm G}]$ 
    & $\sim0.1$
    & $\sim1$
    & $\sim0.01$ 
    & $\sim100$\\
    $\xi_{\rm B}$  
    & $\sim 1$ 
    & $\sim 1$  
    & $\sim1$  
    & $\sim100$ \\
    $\tau_{p\gamma0}$ by Eq.~(\ref{eq:optical_depth_general})
    & $\sim 10^{-5}$ 
    & $\gtrsim 10^{-3}$ 
    & $\sim10^{-4}$ 
    & $\sim1$ \\
    \hline
    $\xi_{\rm UHECR}$: Eq.~(\ref{eq:cr_loading}) 
    & $\sim10-100$ 
    & $\sim10-100$ 
    & $\sim1-10$ 
    & $\sim 1000-10000$\\
    $\xi_{\rm B}$ by acceleration: Eq.~(\ref{eq:hillas_condition}) 
    & $\gtrsim 0.3\eta^2(\frac{Z}{10})^{-2}$ 
    & $\gtrsim 0.3\eta^2(\frac{Z}{1})^{-2}$ 
    & $\gtrsim 1\eta^2(\frac{Z}{10})^{-2}$ 
    & $\gtrsim 0.03\eta^2(\frac{Z}{10})^{-2}$ \\
    $\tau_{p\gamma0}$ by $\nu$ flux: Eq.~(\ref{eq:neutriono_flux_requirement})
    & $\gtrsim 0.3$ 
    & $\gtrsim 0.01$ 
    & $\gtrsim 0.04$
    & $\gtrsim 0.3$ \\
    $\tau_{p\gamma0}$ by escape: Eq.~(\ref{eq:esc_condition}) 
    & $\lesssim 1{(\frac{A}{2Z})}^{4}$ 
    & $\lesssim 1{(\frac{A}{Z})}^{4}$  
    & $\lesssim 1{(\frac{A}{2Z})}^{4}$  
    & $\lesssim 1{(\frac{A}{2Z})}^{4}$ \\
    $\tau_{p\gamma0}$ by nuclei survival: Eq.~(\ref{eq:suvival_condition}) 
    &$\lesssim 0.4(\frac{A}{56})^{-0.21}$ 
    &$\lesssim 0.4(\frac{A}{56})^{-0.21}$ 
    &$\lesssim 0.4(\frac{A}{56})^{-0.21}$ 
    &$\lesssim 0.4(\frac{A}{56})^{-0.21}$ \\
    \end{tabular}
\end{ruledtabular}
\end{table*}

AGNs have been proposed in the literature for many decades as a possible UHECR origin and high-energy neutrino emitters. 
The UHECR emission from the radio-loud (RL) AGN jets
represents a classical scenario discussed in the past and present~\cite{AlvesBatista:2019tlv}. 
The fact that blazars dominate the high-energy gamma-ray sky 
maintains the AGN jets as viable UHECR accelerators~\cite{Murase:2011cy}.
The jet activity in EM emission varies with time, suggesting that 
the UHE particle emission can also be time dependent, 
and neutrinos may be predominantly produced during the flaring phase~\cite{Murase:2018iyl,Yoshida:2022wac}.
However, because the flaring timescale can be the order of months, which is far longer than the real transient class we discuss later, 
we consider the average behavior of the jets here for holding the simplicity.

Table~\ref{table:constraints_AGN} lists the parameters obtained by the unification model for the representative populations/acceleration sites in the AGN family. 
Several remarks on the general constraints should be mentioned here. 
First, larger values of the magnetic field parameter, $\xi_{\rm B}$, are favored by the UHECR acceleration condition (Eq.~(\ref{eq:hillas_condition})), but higher $\xi_{\rm B}$ severely constrains the $p\gamma$ optical depth by the UHECR escape condition (Eq.~(\ref{eq:esc_condition})), in return, which hardly meets the requirement to explain the neutrino flux given by Eq.~(\ref{eq:neutriono_flux_requirement}). 
This is the fundamental reason why the parameter space allowed in the unification scenario is rather limited. This is a solid argument independent of the details on source physics.
Second, the UHECR energetics solely determines the UHECR loading factor condition (Eq.~(\ref{eq:cr_loading})) and any rare class of the object needs $\xi_{\rm UHECR}\gg 1$. This may not be favored by the aspect of the energy budget of the central engine. Also, the source number density has been observationally constrained by the search for small-scale anisotropy in the UHECR arrival direction distribution~\cite{PierreAuger:2013waq} although details depend on the magnetic field due to the time delay of cosmic rays~\cite{Murase:2008sa}.

Jets of BL Lac objects (BL Lacs) are promising sites of UHECR acceleration, especially if UHECRs are nuclei~\cite{Murase:2011cy}. The acceleration condition can be satisfied for $\xi_{\rm B}(\sim 1)$ (e.g., Ref.~\cite{2010MNRAS.402..497G}), and small values of $\tau_{p\gamma0}$ imply that
protons and nuclei can be accelerated to the UHE region without suffering from the energy losses by the photohadronic collisions. However, for this reason, it is typically difficult to consider BL Lacs as the sources that mainly contribute to the all-sky neutrino flux (e.g., Refs.~\cite{Yuan:2019ucv,McDonough:2023ngk}). The estimated optical depth to the photomeson production is
$\displaystyle 
\tau_{p\gamma0}\sim 7\times 10^{-6}\left({B'}/{{\rm 0.1 G}}\right)
\left({\Gamma}/{10}\right)^{-3}
\left({L_\gamma}/{2\times 10^{45}{\rm erg/s}}\right)^{\frac{1}{2}}
\xi_B^{-\frac{1}{2}}$, which indicates that BL Lacs would be too dim to be viable for the unification.

Flat-spectrum radio quasars (FSRQs) can be interesting as the sources of UHECR accelerators and neutrino emitters especially in the presence of external fields~\cite{Murase:2014foa,Dermer:2014vaa,Rodrigues:2017fmu}. Using Eq.~(\ref{eq:optical_depth_general}), the photomeson optical depth due to internal photons is expected to be small, but external photons play an important role as targets because they are boosted in the comoving frame of the jet. Assuming that $L_{\rm ext}\sim0.01L_\gamma$, the optical depth due to external photons is $\displaystyle \tau_{p\gamma}\sim 1\left({B'}/{{\rm 1~G}}\right) \left({\Gamma}/{10}\right)\left({L_{\rm ext}}/{10^{45}{\rm erg/s}}\right)^{\frac{1}{2}}{\xi_B}^{-\frac{1}{2}} \left(\varepsilon_p/\varepsilon_{\rm UHECR}^{\rm FID}\right)^{0.5}$. 
Given that $\tau_{p\gamma0} \sim 0.1-1$, UHECR escape and efficient neutrino production can be concurrent in principle. Large photon luminosities with $L_\gamma \gtrsim 10^{46}\ {\rm erg/s}$ meet the acceleration condition given by Eq.~(\ref{eq:hillas_condition}) even for protons ($Z=1$), while it is challenging for nuclei to survive especially in the presence of external photons. This implies that FSRQs could only be the common origin for UHECR protons and neutrinos, but not for UHECR nuclei. However, we should remark that the unification model by FSRQs would be inconsistent with the other aspects of the neutrino data. Cosmogenic neutrinos produced by UHECR protons emitted from strongly evolved sources such as FSRQs would have overshot the present upper limit of neutrino flux at EeV ($10^{18}\ {\rm eV}$) range placed by IceCube and PAO~\cite{Aartsen:2016ngq,Aab:2019auo}.
The present constraints placed by the cosmogenic neutrino search become certainly
weaker if UHECRs are far dominated by heavy nuclei, but this loophole 
is soon to be substantially narrower when the sensitivity of the neutrino telescopes reaches to $E_\nu^2\Phi_\nu\sim 10^{-9}~{\rm GeV~cm^{-2}~s^{-1}~sr^{-1}}$ by accumulating more data or by the next-generation detectors such as IceCube-Gen2.

UHECRs may be accelerated in weak jets of radio-quiet (RQ) AGN~\cite{Peer:2009vnw}. However, RQ AGNs do not have powerful jets, so that only heavy nuclei can be accelerated to ultrahigh energies. The efficiency of neutrino production is also predicted to be very low, and they are unlikely to contribute to the all-sky neutrino flux without violating gamma-ray data; instead, nuclei may survive against photodisintegration~\cite{Peer:2009vnw}. 
Neutrino production in RQ AGN such as NGC 1068 would preferentially occur in the vicinity of supermassive black holes, including hot coronal regions~\cite{Murase:2015xka,Murase:2022dog}, in which $\tau_{p\gamma}\gtrsim 1$ is expected and the all-sky neutrino flux and neutrino emission from NGC 1068 can simultaneously be explained~\cite{Murase:2019vdl}. However, small radii prevent cosmic rays from being accelerated to ultrahigh energies.

The hot disk of radio-loud AGN (RL AGN) is expected to be radiatively inefficient and collisionless for protons, in which protons could be accelerated without suffering from efficient $p\gamma$ interactions. Protons and nuclei may be accelerated by turbulence via stochastic acceleration~\cite{Kimura:2019yjo,Kimura:2020srn}, and magnetic reconnections could also lead to the UHECR production if the disk is magnetically arrested~\cite{Kimura:2020srn,Kuze:2024sjq}. High-energy neutrinos can also be produced especially in the PeV or higher-energy range. Here we note that the photomeson optical depth in Table~\ref{table:constraints_AGN} should rather be regarded as upper limits, and realistic values considering realistic spectral energy distributions can readily be much smaller~\cite{Kimura:2020srn,Kuze:2024sjq}. 
We also note that large values of $\xi_B$ and $\xi_{\rm UHECR}$ are necessary to satisfy the UHECR energetics, but we note that such a possibility of dark accelerators is energetically allowed when the disk is radiatively inefficient and the accretion power is much larger than the radiation luminosity.

\subsection{Transient sources}

\begin{table*}
    \caption{Same as Table~\ref{table:constraints_AGN} but the various transient objects are listed. Note that parameters have significant uncertainties although only specific parameters are shown in this table. We use $\varepsilon_i^{\rm max}=10^{11}$~GeV. Star-formation evolution is assumed for supernovae and GRBs, while no evolution is assumed for TDEs.}
    \label{table:constraints_transients}
\begin{ruledtabular}
  \begin{tabular}{lcccc}
    & Jetted TDE & TDE wind & LL GRB & Engine-driven SN\\
    \hline
    $\Gamma\beta$ of the outflow 
    & $\sim10$ 
    & $\sim0.3$ 
    & $\sim5$ 
    & $\sim0.3$\\
    Target photon energy & x-ray & opt/UV & x-ray & opt/UV\\
    $L_\gamma\ [{\rm erg/s}]$ 
    & $\sim10^{47}$ 
    & $\sim10^{44}$ 
    & $\sim10^{47}$ & $\sim10^{44}$\\
    $\rho_0 \ [{\rm Mpc^{-3} yr^{-1}}]$ & $\sim10^{-11}-10^{-10}$ & $\sim10^{-7}-10^{-6}$ & $\sim10^{-7}-10^{-6}$ & $\sim10^{-6}-10^{-5}$\\
    $\Delta T \ [{\rm s}]$ 
    & $\sim10^{6}-10^{7}$ 
    & $\sim10^{6}-10^{7}$ 
    & $\sim10^{3}-10^{4}$ 
    & $\sim10^{6}-10^{7}$\\
    \hline
    $R\ [{\rm cm}]$ 
    & ${\rm a~few}\times{10}^{15}$ 
    & $\sim{10}^{17}$ 
    & ${\rm a~few}\times{10}^{15}$ 
    & $\sim{10}^{17}$\\
    $B'\ [{\rm G}]$ 
    & $\sim300$ 
    & $\sim1$ 
    & $\sim100$ 
    & $\sim1$\\
    $\xi_{\rm B}$ 
    & $\sim1$
    & $\sim1$ 
    & $\sim1$ 
    & $\sim1$\\
    $\tau_{p\gamma0}$ by Eq.~(\ref{eq:optical_depth_general}) 
    & $\sim 0.1$
    & $\sim 0.03$ 
    & $\sim 1$
    & $\sim 0.03$ \\
    \hline
    $\xi_{\rm UHECR}$: Eq.~(\ref{eq:cr_loading}) 
    & $\sim100-1000$ 
    & $\sim 1-10$ 
    & $\sim 10-100$ 
    & $\sim 0.1-1$\\
    $\tau_{p\gamma0}$ by $\nu$ flux: Eq.~(\ref{eq:neutriono_flux_requirement})
    & $\gtrsim 0.1$
    & $\gtrsim 0.1$ 
    & $\gtrsim 0.03$ 
    & $\gtrsim 0.03$ \\
    $\xi_{\rm B}$ by acceleration: Eq.~(\ref{eq:hillas_condition}) 
    & $\gtrsim 10^{-2} \eta^2(\frac{Z}{10})^{-2}$ 
    & $\gtrsim 1 (\frac{\eta}{10})^{2}(\frac{Z}{10})^{-2}$ 
    & $\gtrsim 0.01 \eta^2(\frac{Z}{10})^{-2}$ 
    & $\gtrsim 1 (\frac{\eta}{10})^{2}(\frac{Z}{10})^{-2}$ \\
    $\tau_{p\gamma0}$ by escape: Eq.~(\ref{eq:esc_condition}) 
    & $\lesssim 1{(\frac{A}{2Z})}^{4}$  
    & $\lesssim 3{(\frac{A}{2Z})}^{4}$ 
    & $\lesssim 1{(\frac{A}{2Z})}^{4}$ 
    & $\lesssim 3{(\frac{A}{2Z})}^{4}$ \\
    
    $\tau_{p\gamma0}$ by nuclei survival: Eq.~(\ref{eq:suvival_condition}) 
    &$\lesssim 0.4(\frac{A}{56})^{-0.21}$ 
    &$\lesssim 0.4(\frac{A}{56})^{-0.21}$ 
    &$\lesssim 0.4(\frac{A}{56})^{-0.21}$
    &$\lesssim 0.4(\frac{A}{56})^{-0.21}$ \\
    \end{tabular}
\end{ruledtabular}
\end{table*}

Powerful transient objects (see Table~\ref{table:constraints_transients})
have also been discussed in the literature as possible UHECR origins. 

Jetted TDEs are so powerful that protons and nuclei could be accelerated to ultrahigh energies~\cite{Farrar:2008ex,Farrar:2014yla}, and the nucleus survival condition can be satisfied especially if the jet power is not sufficiently large~\cite{Zhang:2017hom}. Neutrino production is reasonably efficient~\cite{Senno:2016bso}, and 
the approximate estimate gives $\displaystyle \tau_{p\gamma}\sim 0.1\left({B'}/{{\rm 300 G}}\right)\left({\Gamma}/{10}\right)^{-3}\left({L_\gamma}/{10^{47}{\rm erg/s}}\right)^{\frac{1}{2}}\xi_B^{-\frac{1}{2}} \left(\varepsilon_p/\varepsilon_{\rm UHECR}^{\rm FID}\right)$.  
One of the caveats in this scenario is that it may be difficult to meet the energetics condition for explaining both neutrinos and UHECRs, and their contribution to the all-sky neutrino flux is expected to be subdominant~\cite{Senno:2016bso}. The requirement for the UHECR loading factor, $\xi_{\rm UHECR}\gtrsim 100$, may also be challenging in light of the efficiency of cosmic-ray acceleration. This shortcoming of the energetics may be resolved if more generous yet-to-be-detected low-luminosity jetted TDEs exist~\cite{Zhang:2017hom} or there may be more missing jetted TDEs. 

Nonjetted TDEs are known to be more abundant objects that are of interest as potential neutrino sources. The disk-driven wind launched from the vicinity of the black hole may be a plausible site for cosmic-ray acceleration~\cite{Zhang:2017hom}. 
The estimated optical depth to photomeson production is $\displaystyle \tau_{p\gamma}\sim 0.05\left({B'}/{{1 \rm G}}\right) \left({L_\gamma}/{10^{44}{\rm erg/s}}\right)^{\frac{1}{2}}  \xi_B^{-\frac{1}{2}}\left(0.3/{\beta}\right) \left(\varepsilon_p/\varepsilon_{\rm UHECR}^{\rm FID}\right)^{0.8}$, 
and $pp$ interactions can also be efficient in this model~\cite{Murase:2020lnu}. It is not trivial to satisfy the acceleration condition, and only nuclei could be accelerated to ultrahigh energies~\cite{Zhang:2017hom}. 
Also, we note that the energy dependence of the $p\gamma$ optical depth is important. In nonjetted TDEs, target photons are mainly thermal, and neutrinos 
are expected to be emitted at PeV or higher energies~\cite{Murase:2020lnu,Winter:2022fpf,Yuan:2023cmd}.  
Because of the nonresonant interactions, $\tau_{p\gamma}$ does not decrease even at higher energies. 

Low-luminosity (LL) GRBs and engine-driven supernovae are among the most promising candidates for the unification model of high-energy neutrinos~\cite{Murase:2006mm,Gupta:2006jm,Kashiyama:2013qet,Murase:2013ffa,Senno:2015tsn} and UHECRs~\cite{Murase:2006mm,Murase:2008mr,Chakraborty:2010pui,Zhang:2017moz,Zhang:2018agl}. 
For LL GRBs, the estimated $p\gamma$ optical depth, $\displaystyle \tau_{p\gamma}\sim 0.4\left({B'}/{{\rm 100 G}}\right)\left({\Gamma}/{5}\right)^{-3}\left({L_\gamma}/{10^{47}{\rm erg/s}}\right)^{\frac{1}{2}}\xi_B^{-\frac{1}{2}}\left(\varepsilon_p/\varepsilon_{\rm UHECR}^{\rm FID}\right)^{1.2}$, 
meets all the conditions listed in Table~\ref{table:constraints_transients}.
Although the rate density, $\rho_0$, and the representative magnetic field, $B'$, are highly uncertain, the LL GRB hypothesis provides the viable parameter space to simultaneously explain the UHECRs and the high-energy neutrinos~\cite{Murase:2006mm,Boncioli:2018lrv}. UHECR acceleration may  preferentially occur at external shocks~\cite{Murase:2008mr,Chakraborty:2010pui}. Since target photons are expected in the optical and ultraviolet range, 
neutrinos are mostly emitted at EeV energies, in which PeV neutrinos may come from different emission regions such as choked jets~\cite{Murase:2013ffa,Senno:2015tsn} (see also Fig.~17 of Ref.~\cite{Zhang:2018agl}).  

Note that we do not consider classical high-luminosity (HL) GRBs as the sources for the unification model. This is because the GRBs have already been excluded as the main sources of IceCube neutrinos through the stacking analysis~\cite{Aartsen:2014aqy,IceCube:2017amx} and observations of the brightest GRB 2201009A~\cite{IceCube:2023rhf,Murase:2022vqf}. However, we note that HL GRBs still remain viable as the dominant sources of UHECRs~\cite{Waxman:1995vg}.

\section{\label{sec:neutrino_xray} Neutrino emission from UHECR sources with x-ray targets}

We have seen that LL GRBs are the promising candidate for the UHE particle emitters
in the unification scheme.
Jetted TDEs and possibly low-luminosity TDEs could be considered. 
Both sources are known as x-ray emitters. This is not an accidental coincidence.
In the photohadronic framework, it is a bulk of x-ray photons with which high-energy cosmic
ray protons are colliding in the production of neutrinos if the bulk Lorentz factor
in plasma is $\Gamma\sim$~3--10, as the condition of the $\Delta$ resonance 
in the photomeson production requires the photon energy being 
$\displaystyle \varepsilon_\gamma\gtrsim (s_\Delta-m_p^2)\Gamma^2/(4\varepsilon_{p})
\sim 15.5(\Gamma/10)^2(\varepsilon_{p}/1 {\rm PeV})^{-1}$~keV. 
Thus, it is naturally expected that
high-energy neutrino emission takes place in an environment
filled with UV to soft x-ray photons. 
These photons, which may be boosted by $\Gamma$,
can be observed as keV-energy x-ray emissions if the sources are optically thin. 
A search for neutrino and soft x-ray coincidence emission is, therefore,  
a powerful approach to identifying such common sources or 
constraining the parameters characterizing the emitters.

In this section, we construct a generic phenomenological model 
to account for the neutrinos observed by IceCube, assuming high-energy protons
colliding with x-ray photons. 
Employing this model, we present how the measured neutrino data
and the searches for the coincident x-ray signals characterize
the high-energy cosmic-ray sources.
The key parameters to characterize the source environment
are the PeV-energy cosmic-ray loading factor $\xi_{\rm CR}$ and the plasma bulk Lorentz factor $\Gamma$.
They are subject to be constrained by the x-ray photon luminosity
measured by x-ray observatories. We demonstrate that the neutrino and soft x-ray
coincidence searches can uniquely probe the parameter space to constrain
the condition for the origin of high-energy neutrinos and UHECRs.
The modeling and the parametrizations are entirely based on 
YM20~\cite{Yoshida:2020div}, but we redefine some of the critical parameters
specifically for lucidly associating the neutrino observations 
to x-ray luminosity observable by the telescopes/sky monitors.

\subsection{\label{subsec:generic_model_xray}
Generic phenomenological model}
%
In YM20~\cite{Yoshida:2020div}, neutrino sources are assumed to 
supply the nonthermal target photons interacting with the accelerated cosmic-ray protons. The reference UHECR proton energy $\varepsilon_{\rm UHECR}^{\rm FID}$ 
was fixed to be 10 PeV in the observer frame, and the photon reference
energy is given by the $\Delta$-resonance
condition as,
\begin{equation}
\varepsilon'_{\gamma0}\approx \frac{(s_\Delta-m_p^2)}{4}
\frac{\Gamma}{\varepsilon_{\rm UHECR}^{\rm FID}}.
\label{eq:photon_energy_ref}
\end{equation}
Thus, the representative photon energy range to calculate the bolometric
photon luminosity $L'_\gamma$, which gauges the central engine power,
must vary from optical to x rays, depending on the bulk Lorentz factor.
In the present parametrization, we instead fix the photon reference energy 
in the plasma rest frame. We denote this energy as $\varepsilon'_{X,b}$.
The target x-ray photon spectrum is represented as a broken power law (BPL),
instead of a single power law employed in YM20, 
and written as
\begin{equation}
\frac{dn_X}{d\varepsilon'_X}= \frac{K'_X}{\varepsilon'_{X,b}}\left(\frac{\varepsilon'_X}{\varepsilon'_{X,b}}\right)^{-\alpha_X},
 \label{eq:target_photon}
\end{equation}
which is distributed from $\varepsilon'_{\rm min}$ to $\varepsilon'_{\rm max}$.
The power-law indices are
\begin{equation}
\alpha_X = \left\{
\begin{array}{l}
\alpha_X^{\rm L}\approx 1 \ \ (\varepsilon'_X <  \varepsilon'_{X,b})
\\
\alpha_X^{\rm H}\approx 2.2 \ \ (\varepsilon'_X\geq \varepsilon'_{X,b})
\\ 
\end{array}
\right.
\label{eq:photon_powerlaw}
\end{equation}
This choice of parameters incorporates a typical x-ray emission
expected in LL GRBs or TDEs.

We introduce the reference bolometric x-ray luminosity $L_X^{\rm REF}$ of the source
defined in the observer frame, which
is subject to the measurement (or constraints) by x-ray observatories with the energy band
of $[\varepsilon_{X,{\rm min}}^{\rm REF}, \varepsilon_{X,{\rm max}}^{\rm REF}]$.
In the case of the Monitor of All-sky x-ray Image ({\it MAXI}) telescope~\cite{2009PASJ...61..999M},
$[\varepsilon_{X,{\rm min}}^{\rm REF}, \varepsilon_{X,{\rm maz}}^{\rm REF}] = [2{\rm keV}, 10{\rm keV}]$
for example. Taking into account the differences in the energy band coverage,
this reference x-ray luminosity $L_X^{\rm REF}$ is connected to the source photon luminosity,
$L_X^{\rm L}$ and $ L_X^{\rm H}$, respectively, which is the bolometric luminosity below 
(above) the BPL break energy $\varepsilon'_{X, b}$. 

The reference x-ray luminosity $L_X^{\rm REF}$ then determines the target photon energy density
and the magnetic field strength $B'$ through the equipartition parameter $\xi_{\rm B}$
as described in YM20, requiring that the magnetic energy density
in the plasma rest frame balances the photon energy density with the spectrum following
the BPL written in Eq.~(\ref{eq:target_photon}) via $\xi_{\rm B}$. 
Note that the photon energy density is given by the entire photon field with luminosity
$L'_X = L_X^{'{\rm L}}+L_X^{'{\rm H}}$ but not $L_X^{\rm REF}$ with the fixed energy band
$[\varepsilon_{X,{\rm min}}^{\rm REF}, \varepsilon_{X,{\rm max}}^{\rm REF}]$.
Note that the shape of the distribution follows the energy spectrum
given by Eq.~(\ref{eq:target_photon}) while the normalization $K'_X$ is solely determined
by $B'$ and $\xi_{\rm B}$ which gives the magnetic energy density balancing to
the photon energy density.

The reference cosmic-ray (CR) proton energy is defined with the $\Delta$-resonance condition
given by Eq.~(\ref{eq:photon_energy_ref}), and obtained by
\begin{equation}
\varepsilon_{p0}[\Gamma]= \frac{(s_\Delta-m_p^2)}{4}\frac{\Gamma}{\varepsilon'_{X,b}},
\label{eq:proton_energy_ref}
\end{equation}
and, therefore, varies with $\Gamma$. For this reference energy, the spectrum of cosmic
ray protons injected from the sources is given by
\begin{equation}
\frac{d\dot{N}_{\rm CR}}{d\varepsilon_p}=\frac{K_{\rm CR}}{\varepsilon_{p0}[\Gamma]}\left(\frac{\varepsilon_p}{\varepsilon_{p0}[\Gamma]}\right)^{-\alpha_{\rm CR}}
\label{eq:UHECRspec}
\end{equation}
and the cosmic-ray bolometric luminosity $L_{\rm CR}$ is
determined by the reference x-ray luminosity $L_X^{\rm REF}$ via
the cosmic-ray loading factor $\xi_{\rm CR}$ as $L_{\rm CR}=\xi_{\rm CR}L_X^{\rm REF}$.
For a more straightforward interpretation of the parameters, we define $L_{\rm CR}$
as the bolometric luminosity above the fixed fiducial 
CR energy $\varepsilon_p^{{\rm FID}}=1$~PeV
which is independent of $\Gamma$. The fiducial cosmic-ray energy is different
from $\varepsilon_{p0}[\Gamma]$ given by Eq.~(\ref{eq:proton_energy_ref}).
We apply the corrections below when necessary. 
The normalization factor of the cosmic-ray spectrum $K_{\rm CR}$ is linked to the x-ray luminosity
as $\displaystyle K_{\rm CR}\approx (\alpha_{\rm CR}-2)\xi_{\rm CR}L_X^{\rm REF}(\varepsilon_{p0}^{\rm FID}/\varepsilon_{p0}[\Gamma])^{\alpha_{\rm CR}-1}/\varepsilon_{p0}^{\rm FID}$ for $\alpha_{\rm CR}> 2$.

Following the formulation in YM20~\cite{Yoshida:2020div},
the optical depth of the photomeson production is given by
\begin{equation}
\tau_{p\gamma}^{\rm L/H} \approx \tau_0^{{\rm L/H}}\left(\frac{\varepsilon_p}{\varepsilon_{p0}[\Gamma]}\right)^{\alpha_X^{\rm L/H}-1}
\label{eq:optical_depth}
\end{equation}
where $\tau_{p\gamma}^{{\rm L}}$ ($\tau_{p\gamma}^{{\rm H}}$)
is the optical depth in the photon field below (above) the break energy $\varepsilon'_{X,b}$.
The cosmic-ray energy independent term $\tau_0^{\rm L/H}$ is written as
\begin{equation}
\tau_0^{\rm L/H} =\sqrt{\frac{L_X^{\rm L/H}}{\xi_{\rm B}^{\rm L/H}}}\frac{B'}{\Gamma^2}
\frac{2}{1+\alpha_X^{\rm L/H}}\frac{\beta^{-1}}{y_{\alpha^{\rm L/H}}(\varepsilon^{'\rm{ L/H}}_{X,{\rm min}},\varepsilon^{'\rm{ L/H}}_{X,{\rm max}})}\mathcal{F},
\label{eq:optical_depth2}
\end{equation}
where $y_\alpha$ is the power-law integral function 
\begin{equation}
y_\alpha(\varepsilon'_{\rm min}, \varepsilon'_{\rm max}) \equiv \int^{\frac{\varepsilon'_{\rm max}}{\varepsilon'_{X,b}}}_{\frac{\varepsilon'_{\rm min}}{\varepsilon'_{X,b}}}
dx x^{-\alpha+1}
\end{equation}
and its integral covers the corresponding energy band below (above) the photon break energy. 
The form factor, $\displaystyle \mathcal{F}\approx 4.3\times 10^{-25}(\varepsilon'_{X,b}/200\ {\rm eV})^{-1}
\ {\rm erg}^{-1}{\rm cm}^{\frac{3}{2}}{\rm s}^{\frac{1}{2}}$
is given by
\begin{equation}
\mathcal{F} = \frac{1}{4\pi \sqrt{c}\varepsilon'_{X,b}}\int ds \frac{\sigma_{p\gamma}(s)}{s-m_p^2}.
\label{eq:form_factor}
\end{equation}

$\xi_{\rm B}^{\rm L/H}$ in Eq.~(\ref{eq:optical_depth2})
is the effective equipartition parameter of the B-field
for the photon energy field below (above) the break energy, and given by
\begin{eqnarray}
    \xi_{\rm B}^{\rm L} & = & \xi_{\rm B}\left(1 + \frac{L_X^{\rm H}}{L_X^{\rm L}}\right) \nonumber\\
    \xi_{\rm B}^{\rm H} & = & \xi_{\rm B}\left(1 + \frac{L_X^{\rm L}}{L_X^{\rm H}}\right)
    \label{eq:effective_B}
\end{eqnarray}



The differential neutrino luminosity emitted from a source
is approximated to be~\cite{Yoshida:2014uka, Yoshida:2020div}

\begin{eqnarray}
\frac{d\dot{N}_{\nu}}{d\varepsilon_{\nu}}\approx  
\int d\varepsilon_p \frac{d\dot{N}_{\rm CR}}{d\varepsilon_p}
(\tau_{p\gamma}^{\rm L}(\varepsilon_p)+\tau_{p\gamma}^{\rm H}(\varepsilon_p))
Y(\varepsilon_{\nu};\varepsilon_p)f_{\rm sup}(\varepsilon_\nu).\,\,\,\,\,\,\,\,\,
\label{eq:general_yield_wz_sync}
\end{eqnarray}

Here  $Y(\varepsilon_{\nu};\varepsilon_p)$ denotes 
the energy distribution of the neutrinos produced
by an interaction of a cosmic-ray proton. The details of the $Y$ formulation
can be found in~\cite{Yoshida:2014uka, Yoshida:2020div}. 
The analytical representation in this formulation is given in Appendix~\ref{apendix:analyticalFormulas}.

The suppression factor in Eq.~(\ref{eq:general_yield_wz_sync}),
$\displaystyle f_{\rm sup}(\varepsilon_\nu)=1-\exp(-t'_{\pi/\mu,{\rm syn}}/t'_{\pi/\mu,{\rm dec}})$,
accounts for synchrotron cooling of
pions and muons. The ratio of the timescale can be rewritten 
in the ratio of the neutrino energy to the critical energy as

\begin{equation}
\frac{t'_{\pi/\mu,{\rm syn}}}{t'_{\pi/\mu,\rm dec}}=\left(\frac{\varepsilon_{\nu,\pi/\mu}^{\rm syn}}{\varepsilon_\nu}\right)^2,
\label{eq:synchrotron_factor}
\end{equation}

and the critical energy is given by

\begin{equation}
\varepsilon_{\nu,\pi/\mu}^{\rm syn} \approx \Gamma\kappa_{\pi,\mu} \sqrt{\frac{6\pi}{\tau_{\pi,\mu}\sigma_TcB'^2}\frac{(m_{\pi/\mu}c^2)^5}{(m_ec^2)^2}},
\label{eq:critical_synchrotron_energy}
\end{equation}

where $\kappa_{\pi,\mu}$ is the inelasticity from pion (muon) to a neutrino in the decay process.
In this work, $\kappa_{\pi}$ is approximated by $\sim 1-r_{\pi}$~\cite{Gaisser:1990vg},
where $r_{\pi} = m_{\mu}^2 / m_{\pi}^2 \simeq 0.57$ is the muon-to-pion mass-squared ratio.
The other fraction goes to a muon, and $\kappa_\mu$ is approximated as $\sim 0.3$.

\begin{table*}
    \caption{Main parameters in the UHECR--neutrino--x-ray unified source model.
    Their baseline values are listed in the third column. The numbers in a bracket are the values 
    for representing the different environments or investigating parameter dependence.
    See the text for the details.
    \label{table:model_parameters}}
\begin{ruledtabular}
  \begin{tabular}{lcc}
  Reference bolometric x-ray luminosity & $L_X^{\rm REF}$&$5\times 10^{46}$($\lesssim 3\times 10^{45}$)[erg/s]\\
  Reference photon energy & $\varepsilon'_{X,b}$ & 0.5 [keV] \\
  X-ray observation band to determine $L_X^{\rm REF}$ & $[\varepsilon^{\rm REF}_{X,{\rm min}},\varepsilon^{\rm REF}_{X,{\rm max}}]$ & [2, 10] [keV]\\
  Source magnetic field &$B'$ & 100 [G] \\
  Equipartition parameter & $\xi_{\rm B}$& 0.1 \\
  Fiducial cosmic-ray energy & $\varepsilon_p^{\rm FID}$ & 1 [PeV] \\
  Fiducial UHECR energy & $\varepsilon_{\rm UHECR}^{\rm FID}$ & 10 [PeV] \\
  Bolometric cosmic-ray luminosity above $\varepsilon_p^{\rm FID}$ & $L_{\rm CR}$ & floating \\
  Cosmic-ray loading factor & $\xi_{\rm CR}$ & $\displaystyle \equiv L_{\rm CR}/L_X^{\rm REF}$ \\
  Cosmic-ray spectrum power-law index & $\alpha_{\rm CR}$ & 2.3 (2.1) \\
  Luminosity density of UHECRs above $\varepsilon_{\rm UHECR}^{\rm FID}$& $Q_{\rm UHECR}$ &  $2.3 (0.85) \times 10^{45}$ [erg Mpc$^{-3}$ yr$^{-1}$] \\
  \end{tabular}
  \end{ruledtabular}
\end{table*}
  
For a given $L_X^{\rm REF}$, the neutrino luminosity $\displaystyle d\dot{N}_\nu/d\varepsilon_\nu$ 
decreases with increasing $\Gamma$.
There are two main reasons for this dependence. 
First, higher $\Gamma$ sources are dimmer
in the plasma rest frame as $L'_X=L_X/\Gamma^2$. Second, the interaction zone becomes more compact
for higher $\Gamma$ due to the Lorentz boost. These factors cause $\sim \Gamma^{-2}$ dependence
on $\tau_0^{\rm L/H}$ in Eq.~(\ref{eq:optical_depth2}) (there is the other minor $\Gamma$
dependence primarily brought by $\varepsilon_{p0}[\Gamma]$).

Table~\ref{table:model_parameters} lists the main parameters and their baseline values
in the model.

\subsection{\label{subsec:diffuse_flux} Contribution to the all-sky neutrino flux}
A population of neutrino sources contributes to the diffuse cosmic background flux.
Assuming emission from standard candles (i.e., identical sources over redshifts), 
the differential flux of diffuse neutrinos from these sources across the universe is calculated by 

\begin{multline}
\Phi_\nu(E_\nu)=\frac{c}{4 \pi}
\int\limits_{z_{\rm min}}^{z_{\rm max}}dz(1+z) \left|\frac{dt}{dz}\right| \\
\times \frac{d\dot{N}_{{\nu_e+\nu_\mu+\nu_\tau}}}{d\varepsilon_\nu}\left|_{\varepsilon_\nu=E_\nu(1+z)}n_0^{\rm eff}\Psi(z), \right.
\label{eq:diffuse_flux}
\end{multline}
where $n_0^{\rm eff}\Psi(z)$ is the comoving source number density given the local source number density
$n_0^{\rm eff}$ and the cosmological evolution factor $\Psi(z)$. 
Sources are distributed between redshift $z_{\rm min}$ and $z_{\rm max}$.

The evolution factor $\Psi(z)$ is parameterized as $(1 + z)^m$ such that the parameter $m$ represents
the ``scale'' of the cosmological evolution that is often used in the literature.
The evolution factor for the neutrino sources is unknown, but
many of the proposed sources are related to supernovae, which approximately trace
the star formation rate (SFR). Following Refs.~\cite{Kotera:2010yn,Yoshida:2012gf}, 
we parametrize $\Psi(z)$ as
\begin{eqnarray}
\Psi(z) \propto \left\{ 
\begin{array}{ll}
(1 + z)^{3.4} & ( 0 \leq z \leq 1 ) \\
{\rm constant} & (1 \leq z \leq 4)
\end{array}
\right. ,
\label{eq:sdf}
\end{eqnarray}
to approximately trace SFR.

\begin{figure}
  \begin{center}
  \includegraphics[width=0.5\textwidth]{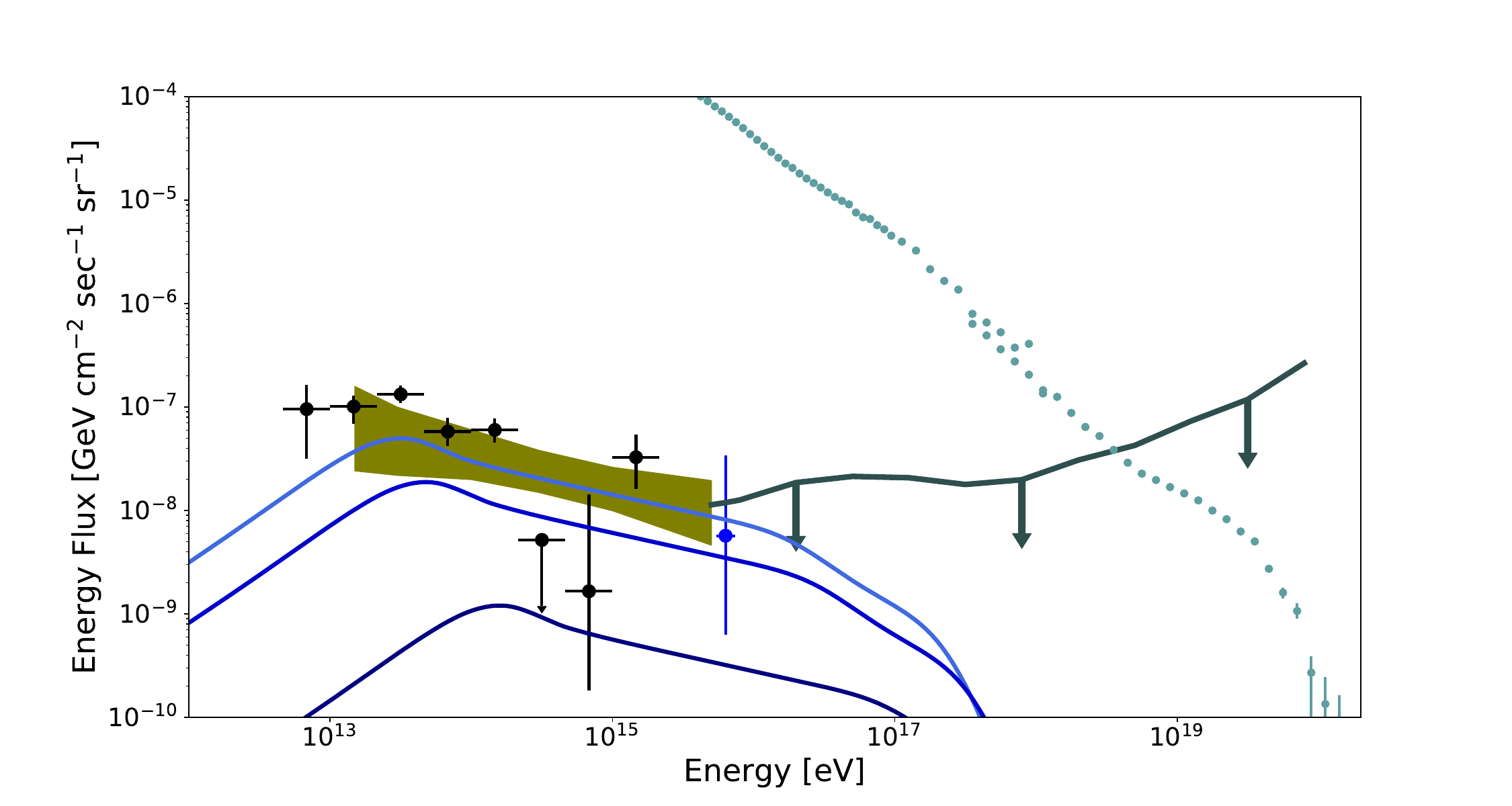}
  \caption{The cosmic diffuse neutrino fluxes from a population of sources with the photon target
  energy density represented by BPL described as Eq.~(\ref{eq:target_photon}),
  along with the neutrino and UHECR observed data shown in Fig.~\ref{fig:energy_fluxes}.
  We assume $L_X^{\rm REF}=5\times 10^{46}$~erg/s, $n_0^{\rm eff}=3\times 10^{-11}\ {\rm Mpc}^{-3}$, 
  $\alpha_{\rm CR}=2.3$,
  and the cosmic-ray loading factor $\xi_{\rm CR}=10$.
  The three cases of $\Gamma$ values, $\Gamma=2$ (top), $\Gamma=3$ (middle), $\Gamma=10$ (bottom)
  are shown.}
  \label{fig:neutrino_diffuse_fluxes}
  \end{center}
\end{figure}
\begin{figure}
  \begin{center}
  \includegraphics[width=0.45\textwidth]{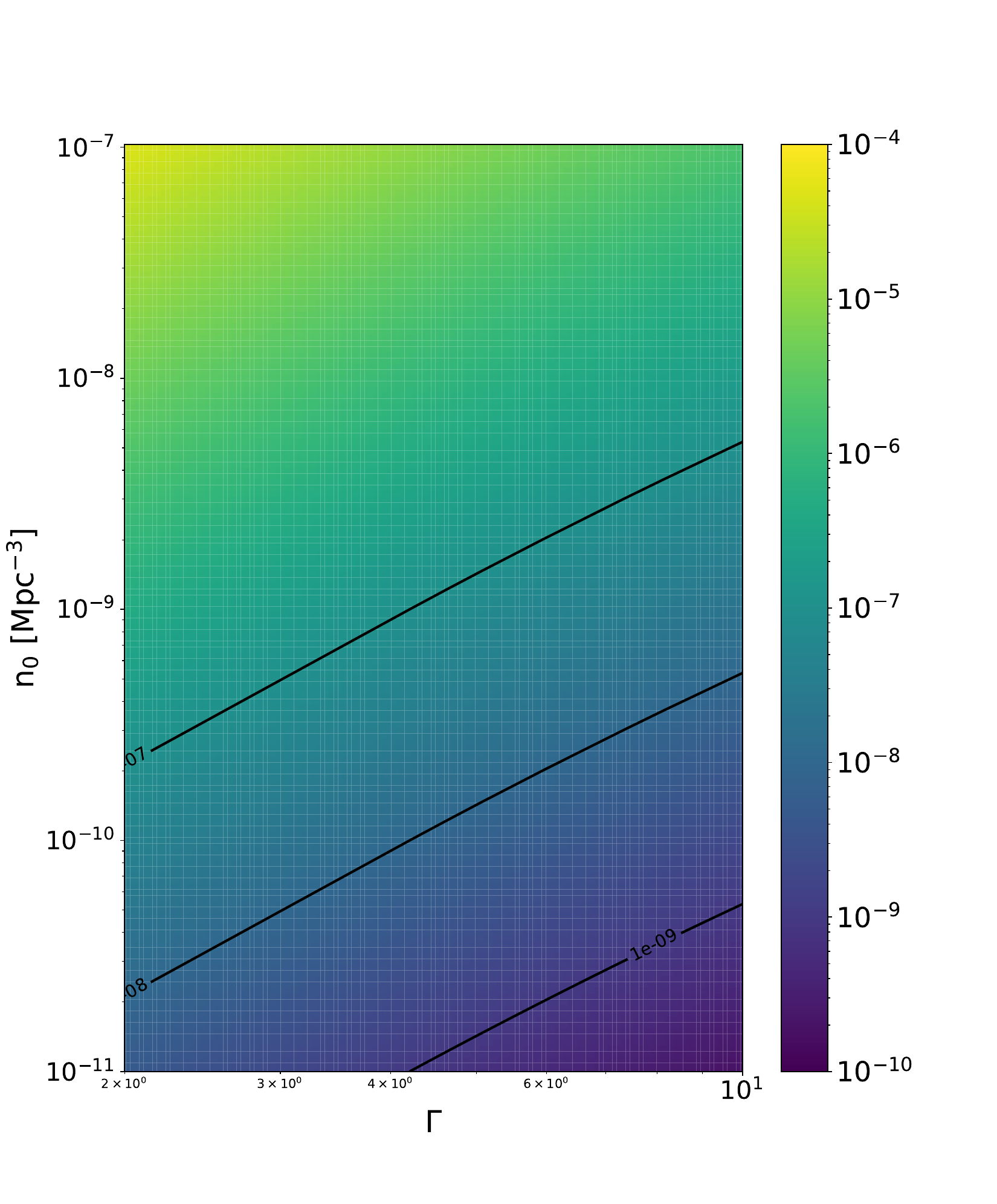}
  \caption{The cosmic diffuse neutrino energy fluxes 
  $E_\nu^2\Phi_{\nu_e+\nu_\mu+\nu_\tau}|_{\rm 1PeV}$
  [${\rm GeV\ cm^{-2}\ s^{-1}\ sr^{-1}}$]
  on the plane of the plasma bulk Lorentz factor $\Gamma$ and the local source number density $n_0^{\rm eff}$.
  We assume $L_X^{\rm REF}=5\times 10^{56}$~erg/s and the cosmic-ray loading factor $\xi_{\rm CR}=10$.}
  \label{fig:neutrino_energy_fluxes2D}
  \end{center}
\end{figure}

Figure~\ref{fig:neutrino_diffuse_fluxes} shows the diffuse background neutrino fluxes produced by
a population of the sources with $L_X^{\rm REF}=5\times 10^{56}$~erg/s, $n_0^{\rm eff}=3\times 10^{-11}\ {\rm Mpc}^{-3}$,
$\alpha_{\rm CR}=2.3$,
and $\xi_{\rm CR}=10$. The IceCube data can constrain these parameters by comparison to the calculated fluxes. 
As shown in Fig.~\ref{fig:neutrino_energy_fluxes2D},
the predicted fluxes in the source parameter plane of $\Gamma$--$n_0^{\rm eff}$
can constrain or estimate the cosmic-ray loading factor $\xi_{\rm CR}$ for a given source
x-ray luminosity $L_X^{\rm REF}$.
We discuss the present constraints in the next section.

\subsection{\label{subsec:constraints} Constraints on source parameters}

The present generic model indicates that the neutrino intensity is determined by
the several source parameters characterizing the photohadronic interaction efficiency
evaluated in the form of optical depth $\tau_0$, and the power of cosmic-ray emission
represented by the cosmic-ray loading factor $\xi_{\rm CR}$.
Since $\displaystyle d\dot{N_{\nu}}/d\varepsilon_\nu \propto \xi_{\rm CR}(L_X^{\rm REF})^{3/2}B'/\sqrt{\xi_{\rm B}}$
($\displaystyle d\dot{N_{\nu}}/d\varepsilon_\nu \propto \xi_{\rm CR}L_X^{\rm REF}$
when the flux reaches the calorimetric limit)
[see Eqs.~(\ref{eq:optical_depth2}), (\ref{eq:neutrino_yield_approx})],
the neutrino data provided by IceCube can constrain these parameters.
Especially if searches for coincident x-ray emission associated with a cosmic neutrino event
determine or constrain $L_X^{\rm REF}$, we can determine or place a bound on $\xi_{\rm CR}$,
a likelihood of a given astronomical object class being the origin of the cosmic rays.
In the following sections, we first present the constraints on the source parameters
with the baseline magnetic field configuration of
$B'=100$~G and $\xi_{\rm B}= 0.1$,
and then expand the constraints to more generic bounds that are valid for the other configurations.

\subsubsection{Constraints from all-sky flux observations}
\begin{figure*}
  \begin{center}
  \includegraphics[width=0.48\textwidth]{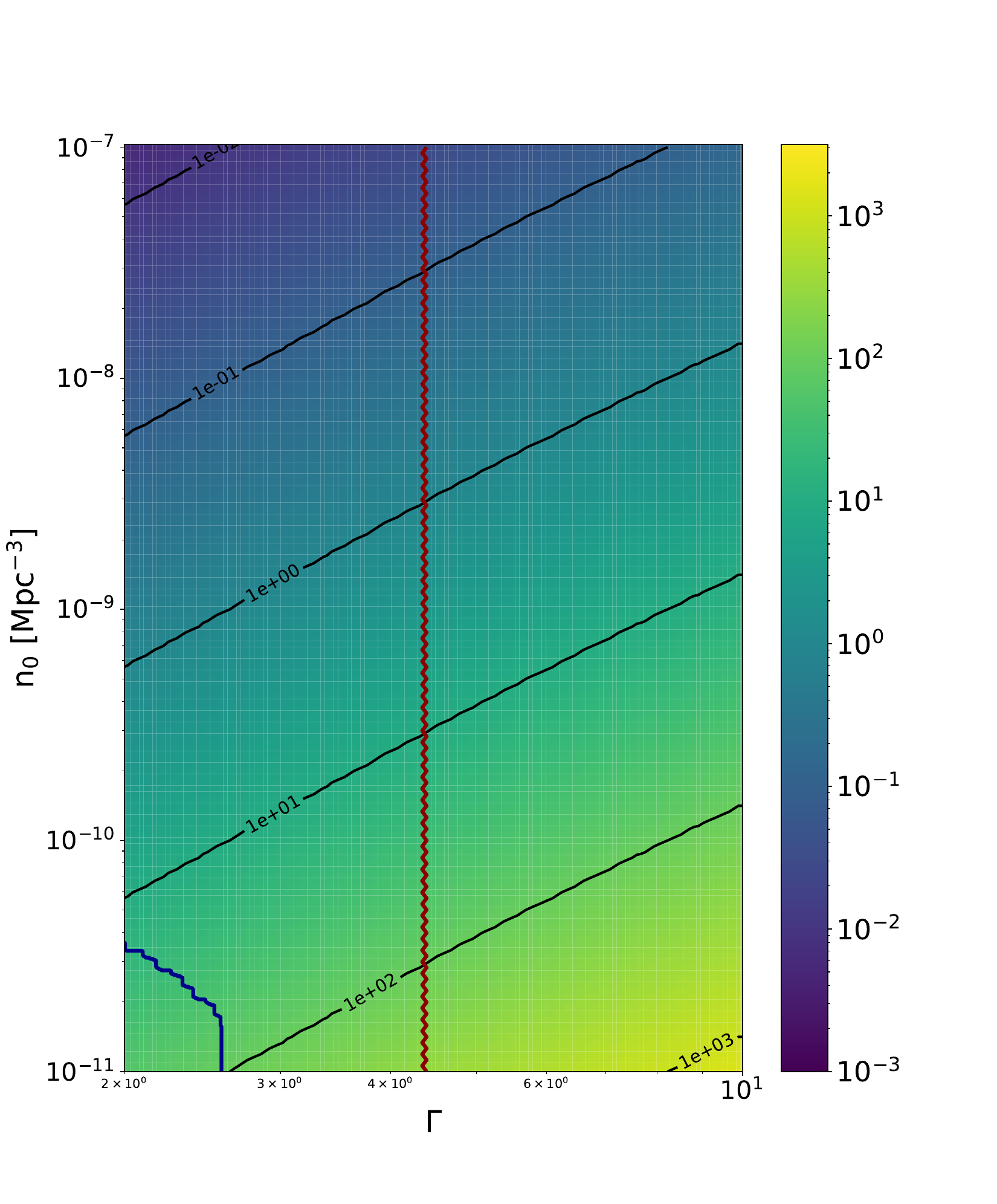}
  \includegraphics[width=0.48\textwidth]{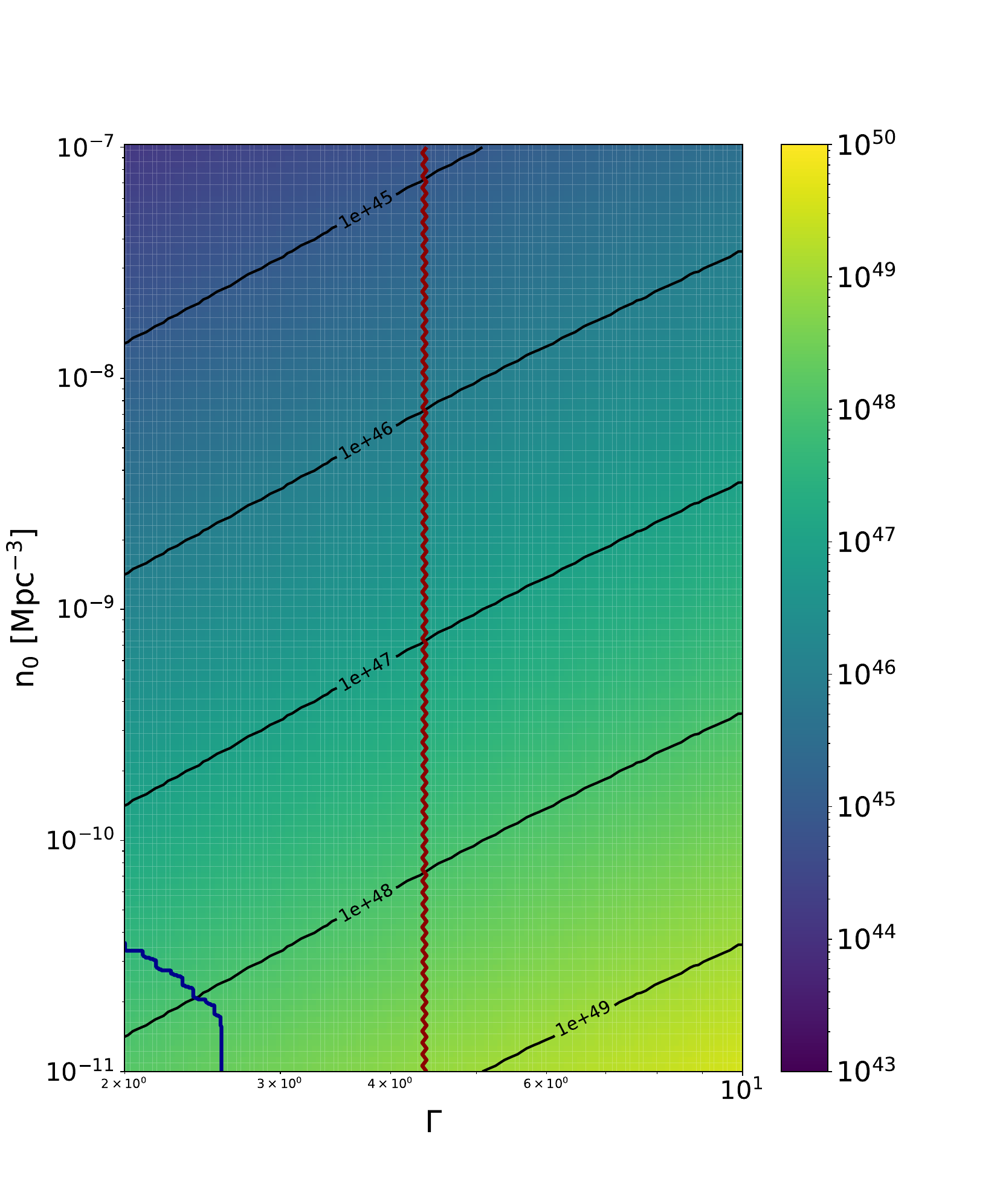}\\
  \vspace{-0.7cm}
  \includegraphics[width=0.48\textwidth]{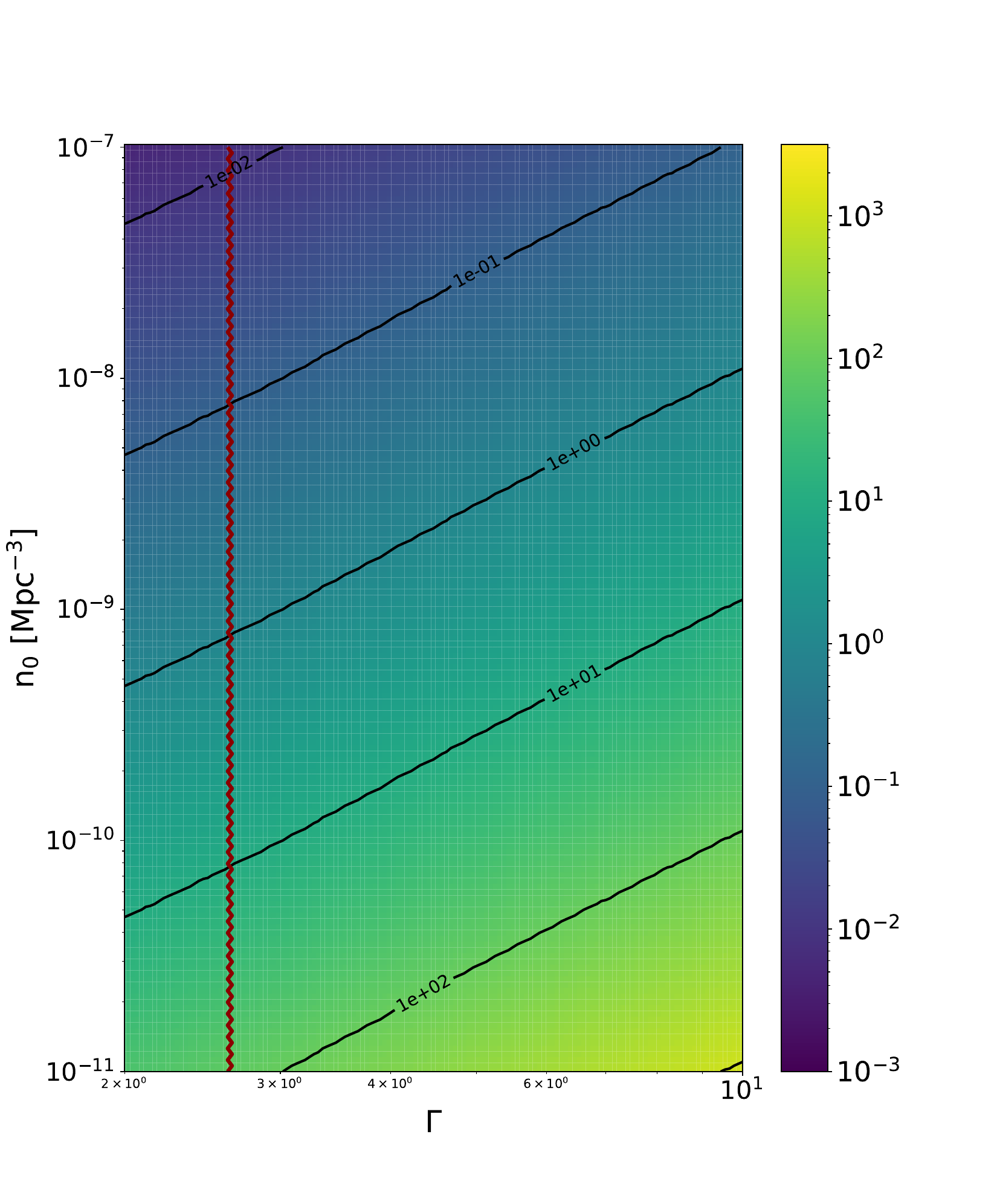}
  \includegraphics[width=0.48\textwidth]{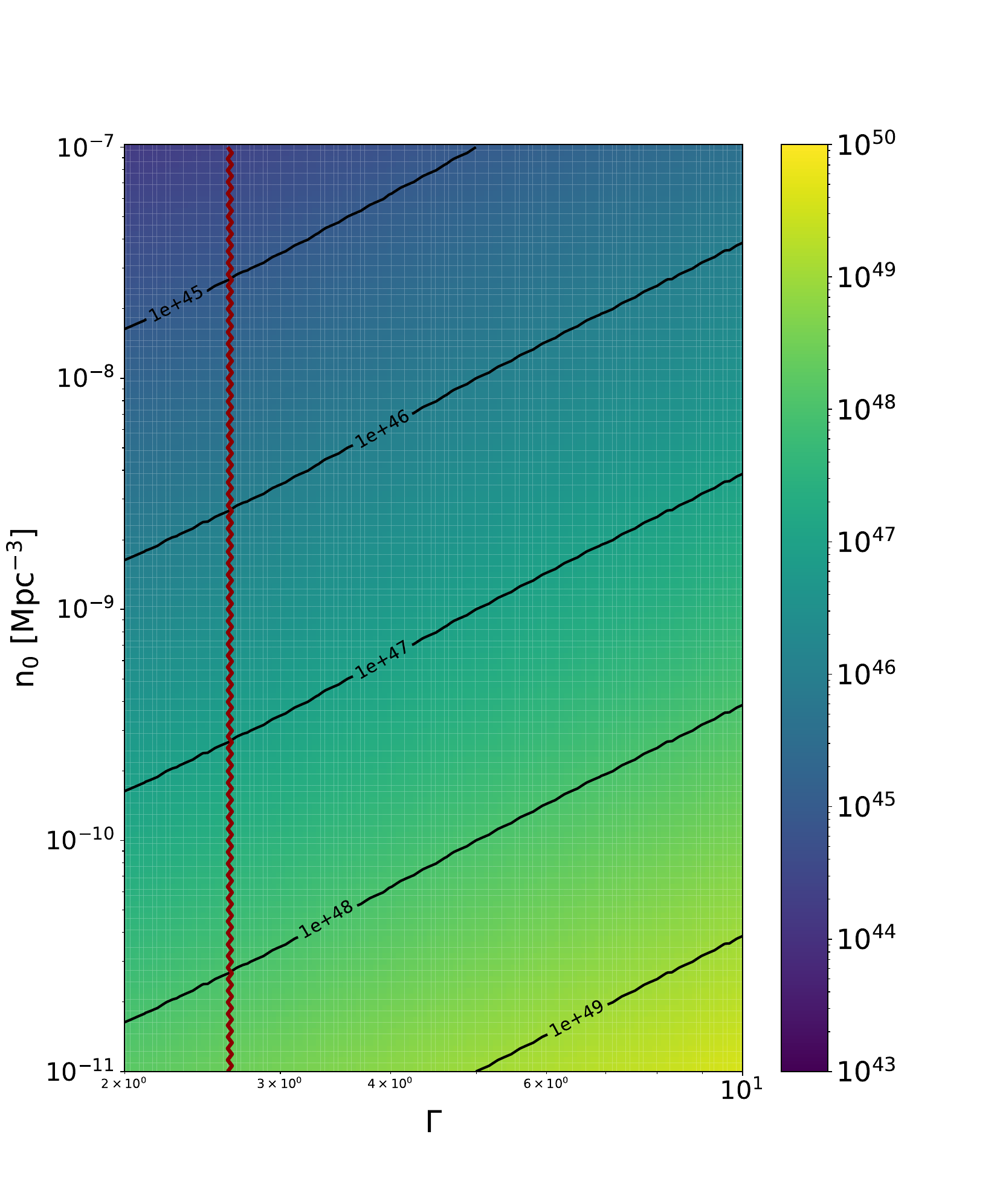}
  \vspace{-0.5cm}
  \caption{The constraints on the cosmic-ray loading factor $\xi_{\rm CR}$ (left)
  and the UHECR bolometric luminosity $L_{\rm UHECR}$ [erg/s]
  above $\varepsilon_{\rm UHECR}^{\rm FID}=10$~PeV (right)
  on the plane of the plasma bulk Lorentz factor $\Gamma$ and the local source number density $n_0^{\rm eff}$.
  The upper (lower) panels display the case of $\alpha_{\rm CR}=2.3 (2.1)$.
  The baseline magnetic field configuration of 
  $B'=100$~G and $\xi_{\rm B}= 0.1$ is assumed.
  The case of the observed x-ray luminosity in the energy band of $[2{\rm keV}, 10{\rm keV}]$
  being $L_X^{\rm REF} = 5\times 10^{46}$~erg/s is shown.
  The parameters to yield the number of neutrino multiplet sources of $2.44$
  per 20 years is shown by the blue curve. The phase space below this curve is disfavored
  if the IceCube measurement with 20-year data does not detect any neutrino multiplet within
  the observational time window of $T_{\rm w}=10^4$~sec.
  The parameters to meet the bolometric luminosity density of cosmic rays above 10 PeV 
  $\mathcal{Q}_{\rm UHECR} = 2.3\times 10^{45} (\alpha_{\rm CR}=2.3), 8.5\times 10^{44} (\alpha_{\rm CR}=2.1)$ 
  erg~Mpc$^{-3}$yr$^{-1}$ is shown by the red line. 
  The parameter space in the right of the red line is unlikely to represent UHECR sources because of the UHECR energetics limit. 
  See the text for details.
  }
  \label{fig:cr_loading_factor2D}
  \end{center}
\end{figure*}

\begin{figure*}
  \begin{center}
  \includegraphics[width=0.48\textwidth]{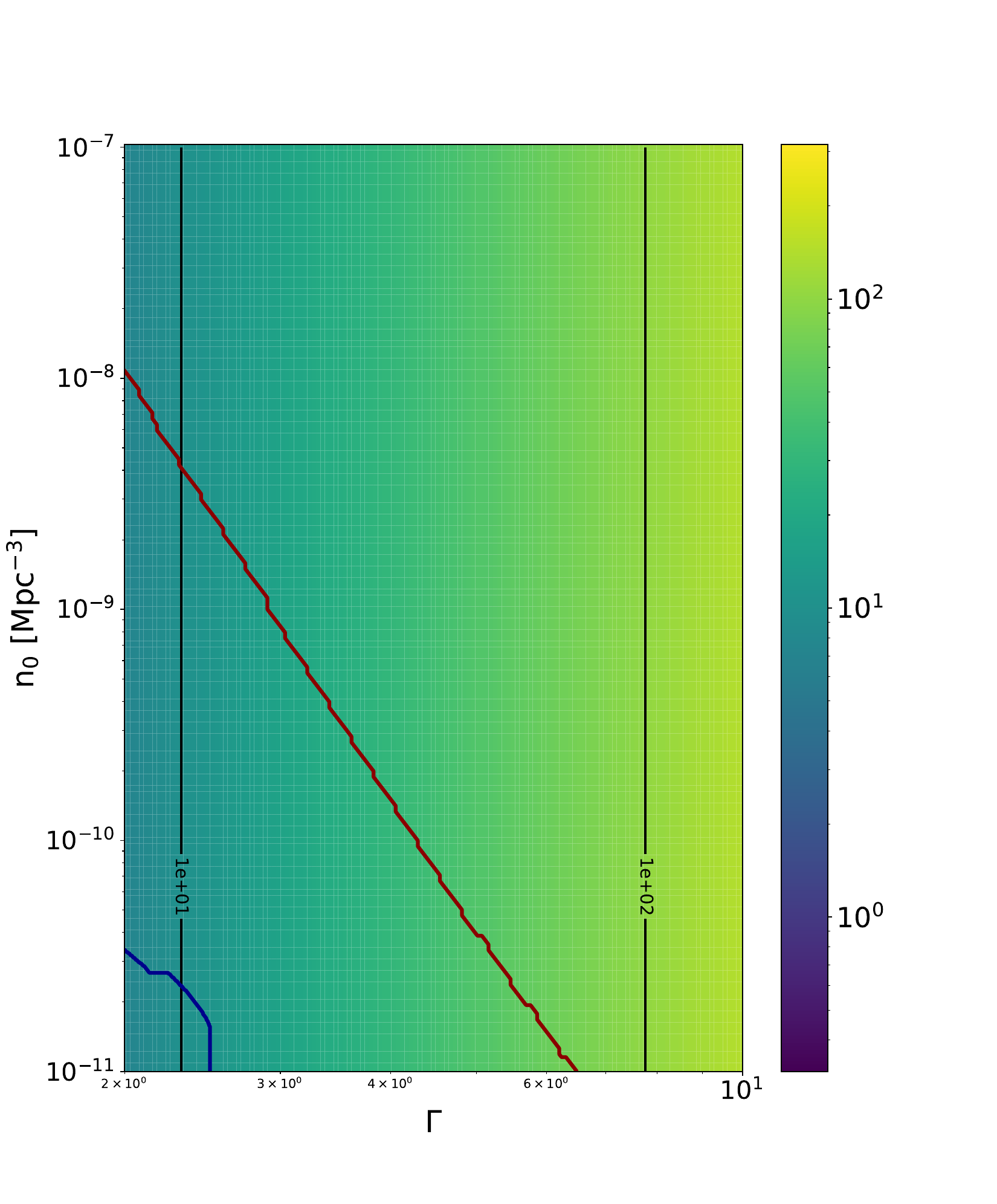}
  \includegraphics[width=0.48\textwidth]{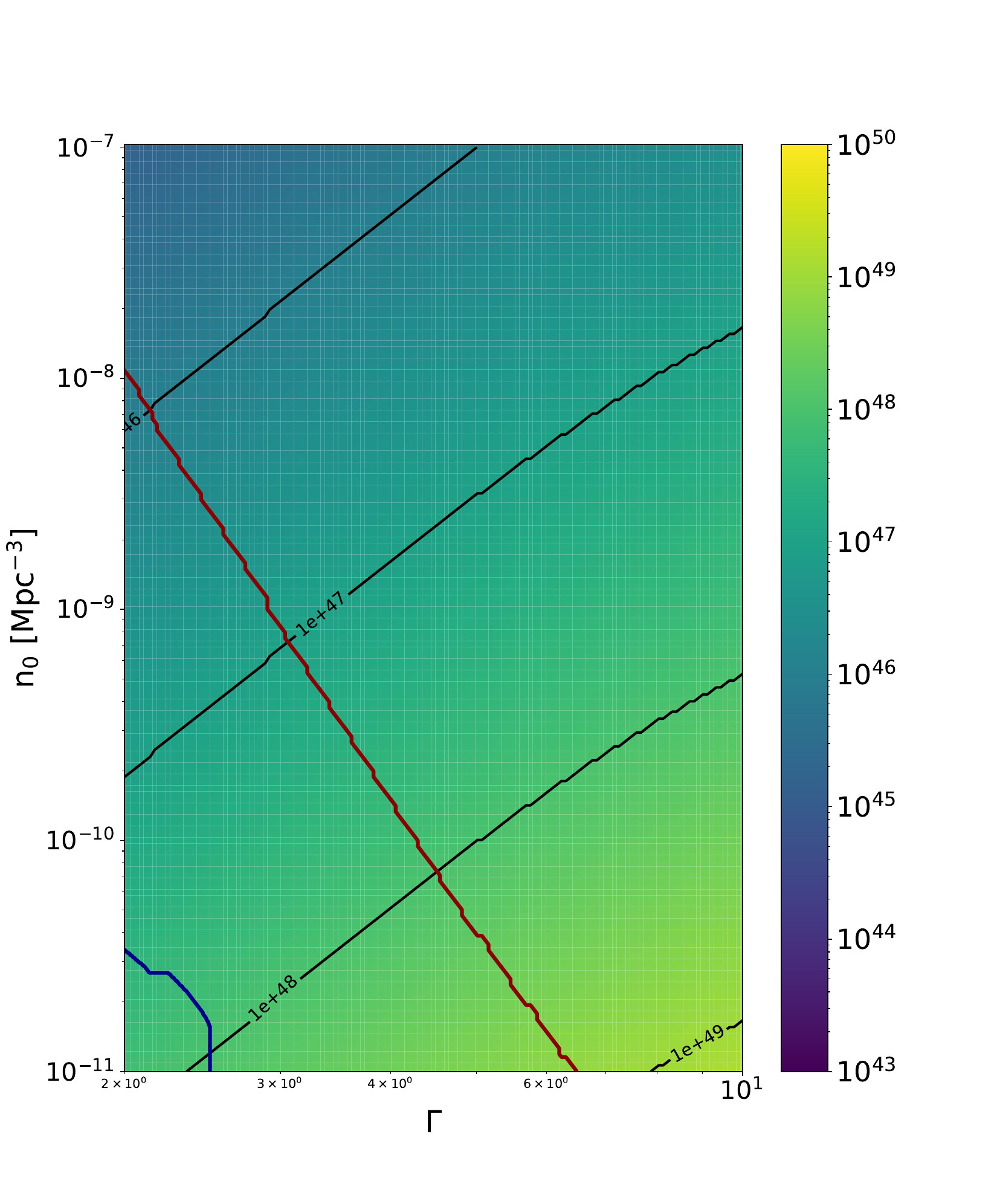}\\
  \vspace{-0.7cm}
  \includegraphics[width=0.48\textwidth]{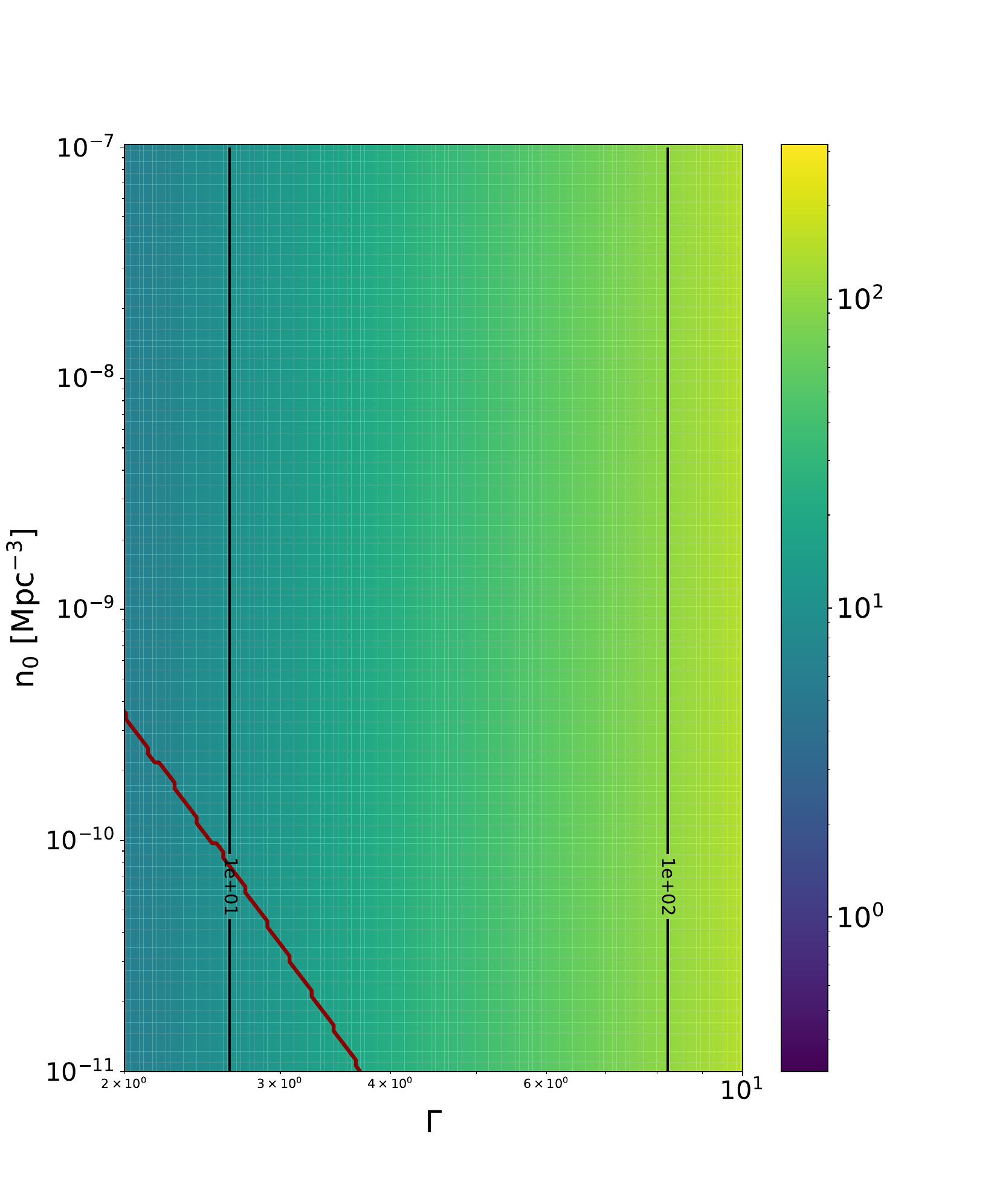}
  \includegraphics[width=0.48\textwidth]{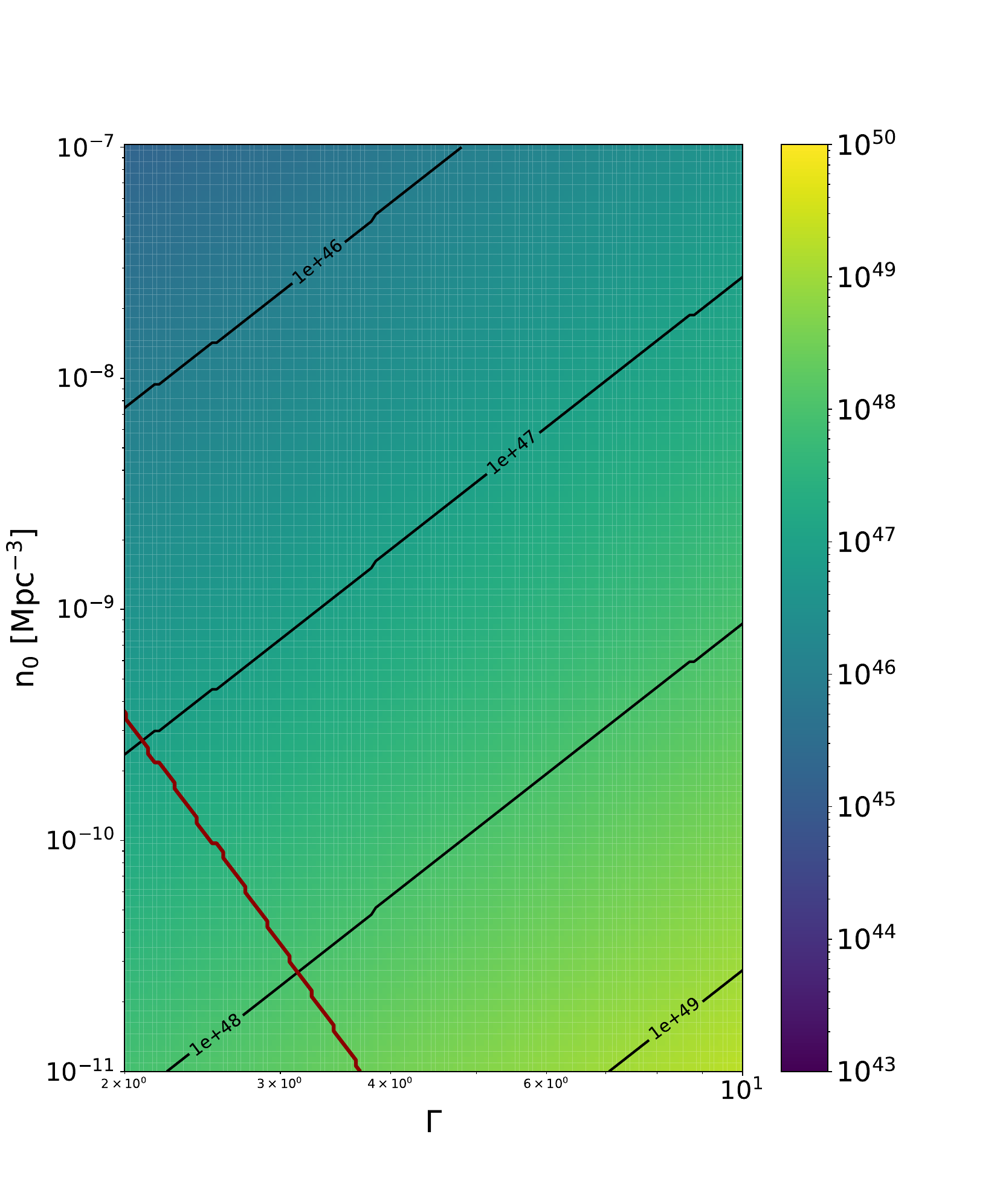}
  \vspace{-0.5cm}
  \caption{The same as Fig.~\ref{fig:cr_loading_factor2D},
  but when the upper limit of the coincident x-ray emission is obtained.
  The upper (lower) panels display the case of $\alpha_{\rm CR}=2.3 (2.1)$.
  The case of $L_X^{\rm UL}=3\times 10^{45}$~erg/s in Eq.~(\ref{eq:ref_Xray_UL}) is displayed.
  The left panel shows the lower bound of cosmic-ray loading factor $\xi_{\rm CR}$
  and the right panel shows the UHECR luminosity $L_{\rm UHECR}$ [erg/s]. 
  The parameter space above the red line is unlikely due to the UHECR energetics constraint.
  The region below the blue curve is disfavored by the neutrino multiplet detection limit. See the text for details.
}
  \label{fig:cr_loading_factor2D_LL}
  \end{center}
\end{figure*}

As the cosmic diffuse neutrino flux $\Phi_\nu(E_\nu) \propto n_0^{\rm eff}\xi_{\rm CR}(L_X^{\rm REF})^{3/2}B'/\sqrt{\xi_{\rm B}}$,
the cosmic-ray loading factor $\xi_{\rm CR}$ is bounded by the comparison of the measured diffuse flux data to
the predicted flux for a given source population characterized by $n_0^{\rm eff}$ and $\Gamma$,
when $L_X^{\rm REF}$ is determined or constrained by the neutrino -- x-ray coincident search.
We consider a standard candle 
source with the photon target energy density given by Eq.~(\ref{eq:target_photon}).
The reference x-ray luminosity $L_X^{\rm REF}$ is calculated in the band of 
$[\varepsilon_{X,{\rm min}}^{\rm REF}, \varepsilon_{X,{\rm max}}^{\rm REF}] = [2{\rm keV}, 10{\rm keV}]$. 
We assume that the population of this standard candle source
with the evolution tracing SFR (Eq.~(\ref{eq:sdf})) gives the cosmic neutrino diffuse background flux.

In order to obtain $\xi_{\rm CR}$ consistent with the IceCube measurements,
we estimated the neutrino intensity
at 100 TeV and the approximated power-law index 
for each of the flux realizations by our model,
and compared them with the profiled likelihood landscape
by the diffuse $\nu_\mu$ data~\cite{IceCube:2021uhz}.  In addition,
we demand the predicted flux to meet the differential $\nu_{e+\mu+\tau}$
flux limit placed by the extremely high-energy (EHE) analysis~\cite{Aartsen:2018vtx}.
This is the same method to extract the allowed parameter space in YM20.

The left panels of Fig.~\ref{fig:cr_loading_factor2D} show
the value of cosmic-ray loading factor $\xi_{\rm CR}$
accounting for the flux data.
The case of $L_X^{\rm REF}=5\times 10^{46}$~erg/s is shown. 
This luminosity is within the expected range of the LL GRBs~\cite{Murase:2006mm}.
We find that lesser populated ($n_0^{\rm eff}\lesssim 10^{-10}\ {\rm Mpc}^{-3}$) source
class with higher plasma Lorentz factor ($\Gamma\gtrsim 5$) requires $\xi_{\rm CR} \gtrsim 100$.
On the contrary, highly populated sources ($n_0^{\rm eff}\gtrsim 10^{-8}\ {\rm Mpc}^{-3}$)
with low $\Gamma$ need $\xi_{\rm CR}\ll 1$. Because we expect $\xi_{\rm CR}$ to be
not far from unity from theoretical viewpoint,
any source classes in these parameter spaces are unlikely to be valid neutrino
source candidates. It is demonstrated that the measurement of x-ray luminosity
leads to bounding the range of the cosmic-ray loading factor, the indicator of likelihood
to be the high-energy neutrino sources for a given set of the source parameters
such as $n_0^{\rm eff}$ and $\Gamma$.

We can extend the bounds obtained here to be universally valid 
for various magnetic field strengths $B'$,
the equipartition parameter $\xi_{\rm B}$ and the x-ray luminosity $L_X^{\rm REF}$. Because 
$\displaystyle \Phi_\nu \propto n_0^{\rm eff}\xi_{\rm CR}(L_X^{\rm REF})^{3/2}B'/(\Gamma^2\sqrt{\xi_{\rm B}}\beta)$
in the most region of the parameter space, 
the constraints on the cosmic-ray loading factor shown in Fig.~\ref{fig:cr_loading_factor2D}, 
which are essentially placed by requiring that the neutrino energy flux does not violate the EHE limit, are approximately represented as 
\begin{multline}
\xi_{\rm CR}\lesssim 15 (11) \left(\frac{L_X^{\rm REF}}{5\times 10^{46}\ {\rm erg/s} } \right)^{-\frac{3}{2}} 
\left(\frac{n_0^{\rm eff}}{1\times 10^{-10}\ {\rm Mpc}^{-3}}\right)^{-1}\\
\times \left( \frac{B'}{100\ {\rm G}}\right)^{-1} \left(\frac{\xi_{\rm B}}{0.1} \right)^{\frac{1}{2}} \left(\frac{\Gamma}{10^{0.5}}\right)^2 \beta,
\label{eq:cr_loading_factor_limit}
\end{multline}
for $\alpha_{\rm CR}=2.3\ (2.1)$.
We confirmed that this scaling law in the relativistic plasma flow case ({\it i.e.,} $\beta=1$) is valid except $\Gamma\lesssim 1.4$
where the neutrino flux reaches the colorimetric limit 
departing from this scaling behavior.

This cosmic-ray loading factor can also be translated to the characteristic UHECR luminosity emitted by the source. Using the EHE limit at 20~PeV, $E_\nu^2\Phi_\nu\approx1.9\times{10}^{-8}~{\rm GeV}~{\rm cm}^{-2}~{\rm s}^{-1}~{\rm sr}^{-1}$, which is also comparable to the all-sky neutrino flux in the PeV range, the bolometric luminosity of UHECR emission 
at $\varepsilon_p\ge \varepsilon_{\rm UHECR}^{\rm FID}$ is given by
\begin{multline}
    L_{\rm UHECR}=\xi_{\rm CR} L_X^{\rm REF}
    \left(\frac{\varepsilon_{\rm UHECR}^{\rm FID}}{\varepsilon_p^{\rm FID}}\right)^{-\alpha_{\rm CR}+2} \\
    \simeq 3.6 (3.8)\times 10^{47}\left(\frac{L_X^{\rm REF}}{5\times 10^{46}\ {\rm erg/s} } \right)^{-\frac{1}{2}} \left(\frac{n_0^{\rm eff}}{10^{-10}\ {\rm Mpc}^{-3}}\right)^{-1}\\
\times \left( \frac{B'}{100\ {\rm G}}\right)^{-1} \left(\frac{\xi_{\rm B}}{0.1} \right)^{\frac{1}{2}} \left(\frac{\Gamma}{10^{0.5}}\right)^2\ {\rm erg/s},
    \label{eq:uhecr_luminosity}
\end{multline}
where $\varepsilon_p^{\rm FID}(={\rm 1 PeV})$ is the fiducial cosmic-ray energy to define
the bolometric cosmic-ray luminosity in the present formulation (see Table~\ref{table:model_parameters}) and $\varepsilon_{\rm UHECR}^{\rm FID}={\rm 10 PeV}$ is the threshold cosmic-ray energy to define the bolometric UHECR luminosity. $L_{\rm UHECR}$ on the plane of $\Gamma-n_0^{\rm eff}$ is shown in the right panels of Fig.~\ref{fig:cr_loading_factor2D}.

\subsubsection{Constraints from searches for x-ray coincidences}

The EM counterparts of neutrinos allow us to search for neutrino sources more efficiently. In this regard, Ref.~\cite{Murase:2016gly} considered limits placed by targeted searches for coincident gamma rays associated with steady neutrino source candidates, assuming $pp$ scenarios (see their Fig.~7). In $p\gamma$ scenarios, hadronic gamma rays are cascaded down to lower energies, and it is better to use target photons as a proxy. The best-known example is HL GRBs, but the gamma-ray search is not effective for many transients such as LL GRBs and TDEs. Here we highlight the importance of all-sky x-ray monitors such as {\it MAXI} and {\it Einstein Probe}.    

If we find no coincident x-ray emission associated with cosmic neutrino signals,
we can set upper limits on $L_X^{\rm REF}$ as a function of $n_0^{\rm eff}$.  
The {\it MAXI} facility~\cite{2009PASJ...61..999M} detected a LL GRB
candidate of $L_X^{\rm REF}\simeq 1.2\times 10^{46}$~erg/s
at a distance of $\sim 360$~Mpc~\cite{Serino:2014wza}. As the detector received $\sim 30$ x-ray photons with the expected background of $\sim 10$ photons in this observation, the indicated sensitivity,
$2\times 10^{-10}~\mathrm{erg}~\mathrm{cm}^{-2}~\mathrm{s}^{-1}$,
will turn into the upper limit on $L_X^{\rm REF}$ for a given $n_0^{\rm eff}$. 
We have
\begin{equation}
    L_X^{\rm REF}\leq L_X^{\rm UL}\left(\frac{n_0^{\rm eff}}{5.2\times 10^{-9}\ {\rm Mpc}^{-3}}\right)^{-\frac{2}{3}},
    \label{eq:ref_Xray_UL}
\end{equation}
$L_X^{\rm UL}=3\times 10^{45}$~erg/s corresponds to the estimated value
for the {\it MAXI} sensitivity, and the Euclidean geometry and no enhancement from the background are assumed. 

On the other hand, when we consider $p\gamma$ scenarios accounting for the all-sky neutrino flux measured by IceCube, we can derive upper limits on the neutrino luminosity or lower limits on $n_0^{\rm eff}$~\cite{Murase:2016gly}. 
This is clear in the calorimetric limit for neutrino production, in which $\displaystyle \Phi_\nu \propto n_0^{\rm eff}\xi_{\rm CR}L_X^{\rm REF}$, and we obtain ``lower limits'' on the cosmic-ray loading factor.  If the system is optically thin to the photomeson production, we have $\displaystyle \Phi_\nu \propto n_0^{\rm eff}\xi_{\rm CR}(L_X^{\rm REF})^{3/2}B'/\sqrt{\xi_{\rm B}}$, and the lower bound becomes independent of $n_0^{\rm eff}$ as $L_X^{\rm REF}\propto {n_0^{\rm eff}}^{-2/3}$. 
Then, assuming that x-ray transients such as LL GRBs and jetted TDEs contribute to the all-sky neutrino flux via $p\gamma$ interactions with the x rays, the upper limit on the x-ray luminosity of neutrino transients leads to the following constraint on the parameter space, 
\begin{multline}
\xi_{\rm CR}\gtrsim 19 (14) \left(\frac{L_X^{\rm UL}}{3\times 10^{45}\ {\rm erg/s} } \right)^{-\frac{3}{2}} \\
\times \left( \frac{B'}{100\ {\rm G}}\right)^{-1} \left(\frac{\xi_{\rm B}}{0.1} \right)^{\frac{1}{2}} \left(\frac{\Gamma}{10^{0.5}}\right)^2\beta,
\label{eq:cr_loading_factor_limit_LL}
\end{multline}
for $\alpha_{\rm CR}=2.3 (2.1)$.

Figure~\ref{fig:cr_loading_factor2D_LL} shows the resulting lower bounds on $\xi_{\rm CR}$. The obtained lower bounds are nearly independent of $n_0^{\rm eff}$ as expected, which are considered as general bounds under the present sensitivity of x-ray detection by a wide field of view observatory such as {\it MAXI}.
We note that both optically thin and thick cases are consistently considered in numerical calculations. 

The corresponding UHECR luminosity needed for explaining the all-sky neutrino flux is given by Eq.~(\ref{eq:uhecr_luminosity}), which is now constrained as
\begin{multline}
    L_{\rm UHECR}=\xi_{\rm CR} L_X^{\rm REF}\left(\frac{\varepsilon_{\rm UHECR}^{\rm FID}}{\varepsilon_p^{\rm FID}}\right)^{-\alpha_{\rm CR}+2} \\
    \gtrsim 4.0 (4.2) \times 10^{47}\left(\frac{L_X^{\rm UL}}{3\times 10^{45}\ {\rm erg/s} } \right)^{-\frac{1}{2}}
    \left(\frac{n_0^{\rm eff}}{10^{-10}\ {\rm Mpc}^{-3}}\right)^{-\frac{2}{3}}\\
\times \left( \frac{B'}{100\ {\rm G}}\right)^{-1} \left(\frac{\xi_{\rm B}}{0.1} \right)^{\frac{1}{2}} \left(\frac{\Gamma}{10^{0.5}}\right)^2\ {\rm erg/s}.
    \label{eq:uhecr_luminosity_UL}
\end{multline}
This lower-limit luminosity is shown in the right panel of Fig.~\ref{fig:cr_loading_factor2D_LL}
for the baseline magnetic field configuration of $B'=100$~G and $\xi_{\rm B}=0.1$.

\subsubsection{Constraints from neutrino multiplet detection}
A population of nearby luminous transient neutrino sources causes detectable neutrino multiplets, two (doublet) or more neutrinos originating from the same direction $\leq \Delta\Omega$ within a certain time frame. A null detection of such neutrino multiplets also constrains the source neutrino energy outputs and the flare(burst) rate density~\cite{Yoshida:2022idr,Murase:2016gly},
as they decide how often we see bright transient neutrino emission from nearby sources. In the present neutrino--x-ray common source modeling, these constraints translate to the bounds on the plasma bulk Lorentz factor $\Gamma$ and the local source number density $n_0^{\rm eff}$ for a given x-ray luminosity $L_X^{\rm REF}$.

The number of neutrino sources that could produce multiplets
from a $2\pi$ solid angle sky detected in the observation time $T_{\rm obs}$
is given by~\cite{Murase:2016gly}
\begin{equation}
  N_{2\pi}^{\rm M}= \frac{2\pi}{4\pi}\frac{T_{\rm obs}}{T_{\rm w}}\int\limits_{z_{\rm min}}^{z_{\rm max}} dz d_z^2(1+z)\left|\frac{dt}{dz}\right| P_{\rm p}^{n\geq 2}[\mu^{\rm s}]n_0^{\rm eff}\psi(z),
  \label{eq:numbe_of_sources_singlet}
\end{equation}
where 
$T_{\rm w}\ll T_{\rm obs}$ is a time window to search for neutrino multiplets, 
$P_{\rm p}^{n\geq 2}$ is the Poisson probability of producing multiple neutrinos 
for the mean number of neutrinos $\mu^{\rm s}$ from a source at redshift $z$, given by
\begin{equation}
  \mu^{\rm s}(\Omega, z) = T_{\rm w}~\frac{1}{3}\int dE_\nu A_{\nu_\mu}(E_\nu, \Omega)\phi_\nu^{\rm fl}(E_\nu, z).
  \label{eq:event_ratefrom_PS}
\end{equation}
Here $A_{\nu_\mu}$ is the detection area of cosmic muon neutrinos for a given neutrino telescope.
Note that the $1/3$ factor is applied to convert the all-flavor-sum neutrino flux to that of per-flavor, 
assuming the equal neutrino flavor ratio. $\phi_\nu^{\rm fl}$ is given by 
\begin{equation}
  \phi_\nu^{\rm fl} = \frac{1}{4\pi d_z^2}\frac{d\dot{N}_{\nu_e+\nu_\mu+\nu_\tau}}{d\varepsilon_\nu}\left|_{\varepsilon_\nu = E_\nu(1+z)} \right.
  \label{eq:point_source_flux}
\end{equation}
where $\varepsilon_\nu$ and 
$E_\nu=\varepsilon_\nu{(1+z)}^{-1}$ are the neutrino energies at the time of emission and arrival at Earth’s surface, respectively.
The proper distance, $d_z$, is calculated via
\begin{equation}
d_z = \frac{c}{H_0}\int\limits_0^{z} dz^{'} 
\frac{1}{\sqrt{\Omega_{\rm M}(1+z^{'})^3+\Omega_\Lambda}}.
  \label{eq:proper_distance_z}
\end{equation}

We consider the possible constraints placed by the neutrino multiplet search, using $A_{\nu_\mu}$ provided by an underground neutrino telescope model with a 1 km$^3$ detection volume 
~\citep{Gonzalez-Garcia:2009bev,Murase:2016gly} as a benchmark. 
The search time window $T_{\rm w}$ is usually chosen to accommodate a typical flare timescale in a transient neutrino source.
Keeping LL GRBs as a baseline source candidate in mind, we set $T_{\rm w}=10^4$ sec to demonstrate how a multiplet search could constrain the source characteristics. Note that if the actual burst time $\Delta T$ is shorter than $T_{\rm w}/(1+z)$, the total neutrino energy output from a source $\varepsilon_\nu^{\rm fl} = L_\nu\Delta T$ is fully observable. The energy contributing to the multiplet detection, $L_\nu T_{\rm w}$, would be less than
$\varepsilon_\nu^{\rm fl}$ otherwise. A similar argument is also held for total cosmic-ray energy output. 
We also point out that the sensitivity to $L_X^{\rm REF}$ by an x-ray telescope is always conservative
when the exposure time of an x-ray monitoring telescope for a given patch of the sky, $T_X$, is
shorter than $T_{\rm w}$. The observation by the {\it MAXI} telescope belongs to this case as $T_X=40$~sec.

For $T_{\rm w}=10^4$~sec, the number of the expected atmospheric neutrino background is $\ll 10^{-2}$ for $\Delta\Omega\approx 1^\circ \times 1^\circ$ in IceCube. 
Therefore, the search for neutrino multiplets can be conducted 
under a background-free environment, and the multiplet source number $N_{2\pi}^{\rm M}$ directly tells the sensitivity. A null detection of any neutrino multiple events by the observation with $T_{\rm obs}=20$~year would constrain the viable parameter space of $\Gamma-n_0^{\rm eff}$ by setting $N_{2\pi}^{\rm M}\leq 2.44$ 
(the Feldman-Cousins 90\% C.L. Poisson upper limit~\cite{Feldman:1997qc}).
The blue curves in Figs.~\ref{fig:cr_loading_factor2D} and \ref{fig:cr_loading_factor2D_LL} represent the resultant constraints. 
The parameter space below the curve is ruled out by 90 \% C.L. The space for rarer source population ({\it i.e.}, smaller $n_0^{\rm eff}$) is disfavored as expected. Note that the multiplet search cannot constrain the parameters by the present IceCube sensitivity in the harder UHECR spectrum case ($\alpha_{\rm CR}=2.1$) because the resultant hard neutrino spectrum yields a lower event rate.

One can notice that the bound placed in the parameter space on the $\Gamma - n_0^{\rm eff}$ plane for the fixed x-ray luminosity case ({\it i.e.,} Fig.~\ref{fig:cr_loading_factor2D}) is nearly identical to that for the upper limit case ({\it i.e.,} Fig.~\ref{fig:cr_loading_factor2D_LL}), indicating that the constraints on $n_0^{\rm eff}$ and $\Gamma$ are independent of $L_X^{\rm REF}$.
We demonstrate that this is indeed true, and the bound is universal regardless of not only the x-ray luminosity but also the magnetic field configuration of $B'$ and $\xi_{\rm B}$ by the following argument.
$\displaystyle P_{\rm p}^{n\geq 2}\simeq (\mu^{\rm s})^2/2$ as $\mu_s\ll 1$.
Then $\displaystyle N_{2\pi}^{\rm M}\propto n_0^{\rm eff}\xi_{\rm CR}^2(L_X^{\rm REF})^3B'^2/(\xi_{\rm B}d_{z^{\rm min}})$.
As the proper distance of a source at $z=z_{\rm min}$, $d_{z^{\rm min}}$, is approximately $\displaystyle \sim \left(3/4\pi n_0^{\rm eff}\right)^{1/3}$, $\displaystyle N_{2\pi}^{\rm M}\propto {n_0^{\rm eff}}^{4/3}\xi_{\rm CR}^2(L_X^{\rm REF})^3B'^2/\xi_{\rm B}$.
Then the cosmic-ray loading factor upper limit set by $N_{2\pi}^{\rm M}\leq 2.44$ is written as $\xi_{\rm CR}^{\rm UL}\propto {n_0^{\rm eff}}^{-2/3}(L_X^{\rm REF})^{-3/2}\sqrt{\xi_{\rm B}}/B'$.
The cosmic-ray loading factor determined by the cosmic diffuse neutrino flux must be lower than this upper limit, {\it i.e.,} $\xi_{\rm CR}\leq \xi_{\rm CR}^{\rm UL}$. As represented in Eq.~(\ref{eq:cr_loading_factor_limit}), $\xi_{\rm CR}\propto {n_0^{\rm eff}}^{-1}(L_X^{\rm REF})^{-3/2}\sqrt{\xi_{\rm B}}/B'$ [primarily because  $\Phi_\nu(E_\nu) \propto n_0^{\rm eff}\xi_{\rm CR}(L_X^{\rm REF})^{3/2}B'/\sqrt{\xi_{\rm B}}$] which has the same dependencies on $L_X^{\rm REF}, B'$, and $\xi_{\rm B}$ as $\xi_{\rm CR}^{\rm UL}$ follows. Thus, these dependencies are canceled out in the condition of $\xi_{\rm CR}\leq \xi_{\rm CR}^{\rm UL}$.

\subsubsection{Additional constraints for unification scenarios}

\begin{figure*}
  \begin{center}
  \includegraphics[width=0.32\textwidth]{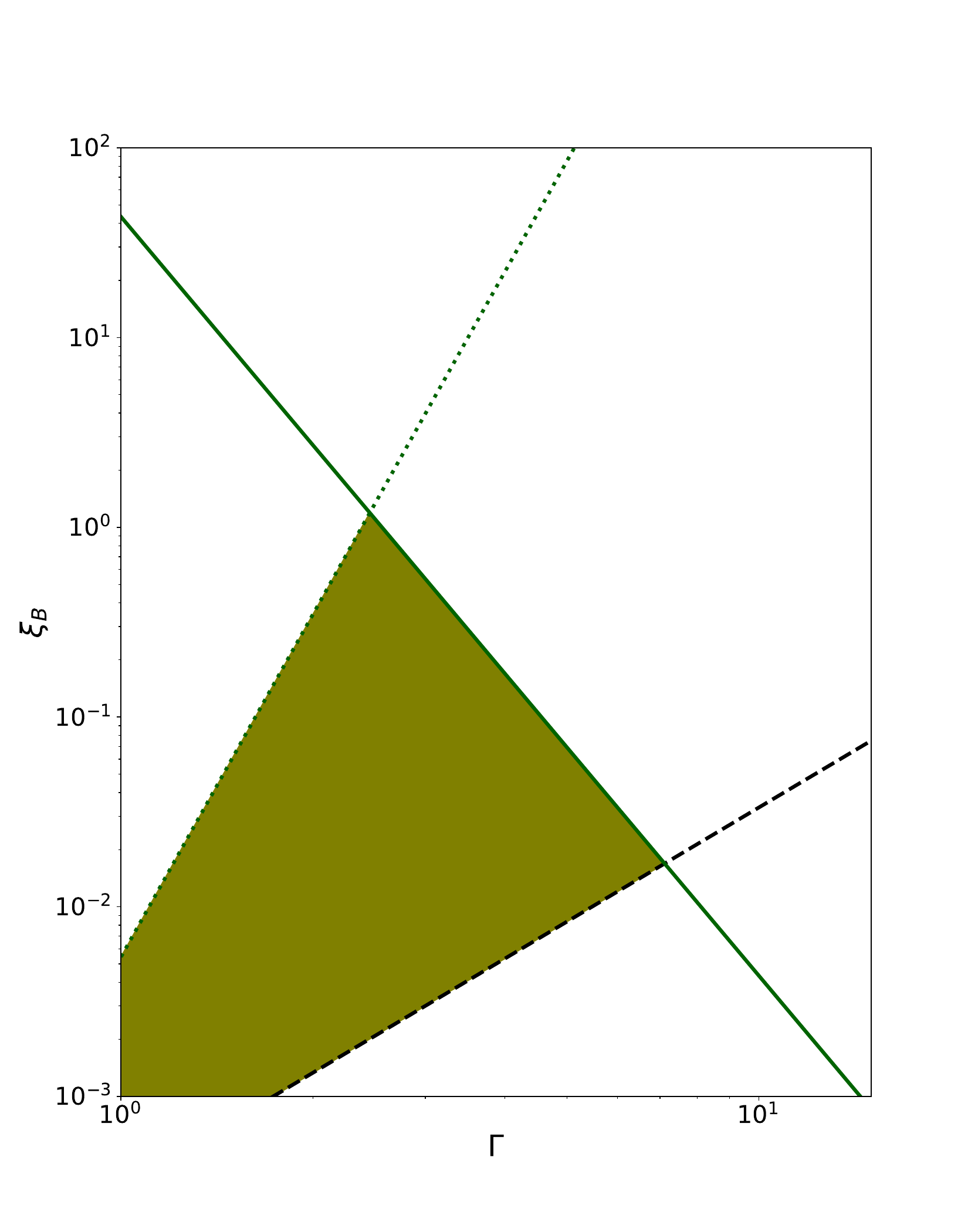}
  \includegraphics[width=0.32\textwidth]{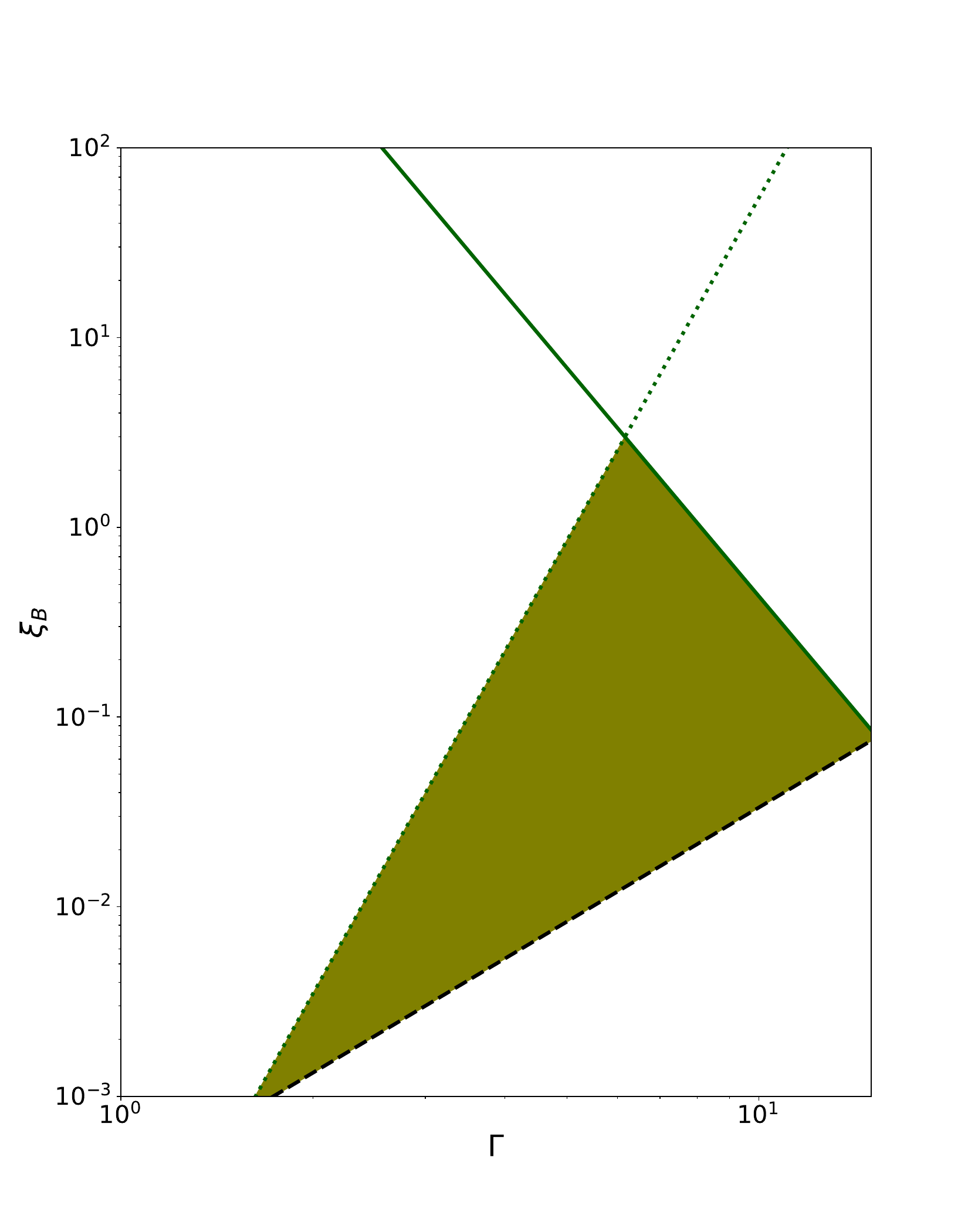}
  \includegraphics[width=0.32\textwidth]{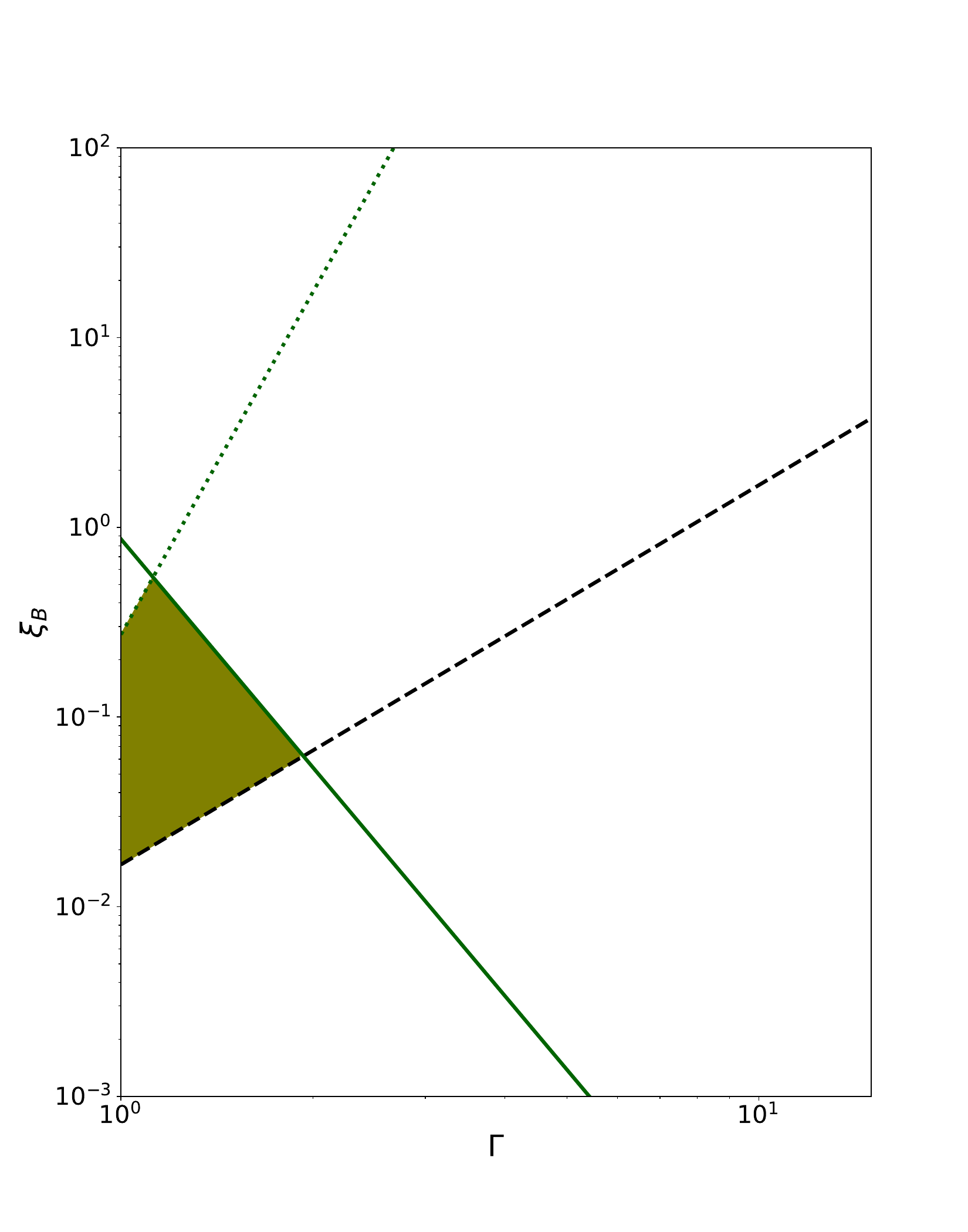}
  \caption{Constraints on the UHECR-neutrino common source model. The colored area
  represents the allowed parameter space by the UHECR energetic condition (thick solid line),
  the acceleration condition (thick dashed line), the escape condition (dotted line),
  and the constraint by multiplet (vertical dash-dotted line).
  Left: The case of $L_X^{\rm REF}=5\times 10^{46}$~erg/s and $B'= 100$~G.
  Middle: The case of $L_X^{\rm REF}=5\times 10^{46}$~erg/s and $B'= 1000$~G.
  Right: The case of $L_X^{\rm REF}=1\times 10^{45}$~erg/s and $B'= 100$~G.
  }
  \label{fig:constraints_xiB_UHECR}
  \end{center}
\end{figure*}

If the neutrino and x-ray emitters we consider here are also emitting UHECRs,
we can place further constraints in the parameter space.
We first discuss the energetics conditions as we did in Sec.~\ref{sec:generic_model}.
The UHECR luminosity density condition demanded by the UHECR energetics
is approximately given by
\begin{equation}
    n_0L_{\rm UHECR} = n_0\xi_{\rm CR}L_X^{\rm REF}\left(\frac{\varepsilon_{\rm UHECR}^{\rm FID}}{\varepsilon_p^{\rm FID}}\right)^{-\alpha_{\rm CR}+2} 
    \lesssim Q_{\rm UHECR}.
    \label{eq:uhecr_energetics_condition}
\end{equation}
$\mathcal{Q}_{\rm UHECR}$~\cite{Yoshida:2020div, Murase:2018utn} is given by Eq.~(\ref{eq:UHECR_energetics})
and listed in Table~\ref{table:model_parameters}.
Because the cosmic-ray loading factor $\xi_{\rm CR}$ or equivalently the UHECR luminosity $L_{\rm UHECR}$
are determined by the cosmic background neutrino flux, 
this condition above sets bounds
on the $\Gamma - n_0$ plane.
The red lines in Figs.~\ref{fig:cr_loading_factor2D} and \ref{fig:cr_loading_factor2D_LL} correspond to
this boundary. Any sources found in the parameter space beyond this line
would overproduce the observed UHECR flux and thus are unlikely to be 
the common UHECR and neutrino origin.
The constraints for the cases other than the baseline magnetic field parameters
of $B'= 100$~G and $\xi_{\rm B}= 0.1$
are approximately obtained by
Eq.~(\ref{eq:uhecr_luminosity}) or (\ref{eq:uhecr_luminosity_UL}) with
Eq.~(\ref{eq:uhecr_energetics_condition}). We get
\begin{multline}
    \left(\frac{L_X^{\rm REF}}{5\times 10^{46}\ {\rm erg/s}}\right)^{-\frac{1}{2}}
\left( \frac{B'}{100\ {\rm G}}\right)^{-1} \\
\times \left(\frac{\xi_{\rm B}}{0.1} \right)^{\frac{1}{2}}\left(\frac{\Gamma}{10^{0.5}}\right)^2 \beta \lesssim 2.0 (1.0),
\label{eq:uhecr_energetics_baseline}
\end{multline}
for the fixed x-ray luminosity case when the UHECR spectrum power-law index $\alpha_{\rm CR}=2.3 (2.1)$ and
\begin{multline}
\left(\frac{n_0^{\rm eff}}{1\times 10^{-10}\ {\rm Mpc}^{-3}}\right)^{\frac{1}{3}}
    \left(\frac{L_X^{\rm UL}}{3\times 10^{45}\ {\rm erg/s}}\right)^{-\frac{1}{2}}\\
\times \left( \frac{B'}{100\ {\rm G}}\right)^{-1} \left(\frac{\xi_{\rm B}}{0.1} \right)^{\frac{1}{2}} \left(\frac{\Gamma}{10^{0.5}}\right)^2 \beta \lesssim 1.8 (0.9),
\label{eq:uhecr_energetics_UL}
\end{multline}
for the scenario when obtaining the upper limit of the associated x-ray luminosity.

UHECR emissions also require
cosmic rays to leave the sources before losing their energies via synchrotron cooling.
This ``survival'' condition leads to the bound on $B'$ and $\Gamma$ represented by~\cite{Yoshida:2020div}
\begin{eqnarray}
B'&\leq& \frac{6\pi m_p^4 c^{9/2}}{\sigma_T m_e^2 {(2\xi_B 
L_X^{\rm REF})}^{1/2}}\frac{\Gamma^3}{\varepsilon_p^{\rm max}} \nonumber \\
&\lesssim& 7\times 10^2 \left(\frac{\Gamma}{10^{0.5}}\right)^3
 \left(\frac{L_X^{\rm REF}}{5\times 10^{46}\ {\rm erg/s}}\right)^{-\frac{1}{2}}\nonumber \\
&& \times \left(\frac{\xi_{\rm B}}{0.1} \right)^{-\frac{1}{2}} 
 \left(\frac{\varepsilon_p^{\rm max}}{10^{10}\ {\rm GeV}}\right)^{-1}\ {\rm G}.
 \label{eq:esc_condition_Lx}
\end{eqnarray}
We note that details depend on the diffusion process, which is particularly relevant for nonrelativistic systems. Here, we conservatively demand protons with energies of 
$\varepsilon_p\lesssim\varepsilon_p^{\rm max}=10^{10}$~GeV to meet the escape condition
since cosmic-ray nuclei would then meet the condition even at energies beyond $10^{11}$~GeV.
The bound can be translated into the upper bound of $\xi_{\rm B}$ for a given $B'$
and we get
\begin{equation}
    \xi_{\rm B} \lesssim 6 \left(\frac{L_X^{\rm REF}}{5\times 10^{46}\ {\rm erg/s}}\right)\left(\frac{B'}{100 {\rm G}}\right)^{2}
    \left(\frac{\Gamma}{10^{0.5}} \right)^{6} 
 \label{eq:esc_condition_Gamma}
\end{equation}

Note that we cannot constrain the parameters by the escape condition when
only the upper limit of the x-ray luminosity is obtained.

The UHECR acceleration condition given by Eq.~(\ref{eq:hillas_condition})
places the lower limit of $\xi_{\rm B}$ for a given x-ray luminosity
as
\begin{equation}
    \xi_{\rm B}\gtrsim 3\times 10^{-3} \left(\frac{L_X^{\rm REF}}{5\times 10^{46}\ {\rm erg/s}}\right)^{-1}
    \left(\frac{\Gamma}{10^{0.5}}\right)^2\left(\frac{\varepsilon_p^{\rm max}}{10^{10}\ \rm{GeV}}\right)^2
\label{eq:xiB_hillas}
\end{equation}

The UHECR energetics condition by Eq.~(\ref{eq:uhecr_energetics_baseline}),
the escape condition by Eq.~(\ref{eq:esc_condition_Gamma}), and
the acceleration condition by Eq.~(\ref{eq:xiB_hillas})
lead to the $n_0^{\rm eff}$ independent bound on the equipartition parameter $\xi_{\rm B}$
and the plasma Lorentz bulk factor $\Gamma$.
Figure~\ref{fig:constraints_xiB_UHECR} shows the resultant allowed parameter space
on $\xi_{\rm B}$ and $\Gamma$.
The energetics and escape conditions lead to $\xi_{\rm B}\lesssim 1$, which is consistent with
Ref.~\cite{Yoshida:2020div}.
Combining the UHECR energetics condition (Eq.~(\ref{eq:uhecr_energetics_baseline}))
and the acceleration requirement (Eq.~(\ref{eq:xiB_hillas})) leads to the universal upper limit 
of the plasma Lorentz bulk factor $\Gamma$ as
\begin{equation}
    \Gamma \lesssim 7 (6) \left(\frac{L_X^{\rm REF}}{5\times 10^{46}\ {\rm erg/s}}\right)^{\frac{1}{3}}
    \left(\frac{B'}{100 {\rm G}}\right)^{\frac{1}{3}},
    \label{eq:Gamma_UL}
\end{equation}
for $\alpha_{\rm CR}=2.3 (2.1)$.


\section{Discussion \label{sec:discussion}}

The synergy between neutrinos and x-ray observations is crucial for testing some models such as LL GRBs. The Wide-Field Telescope (WFT) of {\it Einstein Probe} has a sensitivity of $\sim3\times{10}^{-11}~{\rm erg}~{\rm cm}^{-2}~{\rm s}^{-1}$ at $0.5-4$~keV. For example, for a LL GRB with $L_\gamma\sim{10}^{46}~{\rm erg}~{\rm s}^{-1}$ in the WFT band, the detection horizon is $z_{\rm lim}\sim 0.3-0.4$, and the number of LL GRBs within 1~sr could be $\sim100$ per year, although in reality it would be optimistic since, e.g., the peak energy will not always be in the WFT band. Most of the all-sky neutrino flux comes from LL GRBs with redshifts of $\sim1-2$, which are beyond the detection horizon. Thus, x-ray counterparts should be found for only a fraction of the alerts. Nevertheless, should no coincident neutrinos be discovered in conjunction with a larger sample of x-ray transients identified by {\it Einstein Probe}, the surveyed range of $L_{X}^{\rm REF}$ can be enhanced by a factor of $\sim10$, thereby enabling constraints on $\xi_{\rm CR}$ to be stringent 
as $\gtrsim1000$.

We also note that the constraints from searches for x-ray coincidences and multiplets are general, which are applicable even if neutrinos and UHECRs are produced in different regions (e.g., neutrinos from choked jets and UHECRs from the external reverse shock) but by the same population~\cite{Zhang:2018agl}. 
On the other hand, it is crucial to know the nature of x-ray transients, including source redshifts. Some of the x-ray transients can be HL GRBs at high redshifts~\cite{Liu:2024ejj,Levan:2024bou,Gillanders:2024aoi}, in which we will be able to constrain rarer and more luminous gamma-ray transients. 

In addition to real-time searches for neutrinos, stacking analyses will also be useful. These analyses allow us to look for lower-energy neutrinos, although the current sensitivity is largely limited by the sample size. Performing stacking analyses with optical transients~\cite{Senno:2017vtd,Esmaili:2018wnv,Chang:2022hqj,IceCube:2023esf} 
and follow-up searches for optical counterparts ~\cite{Yoshida:2022idr} will also be useful, especially in the era of the Rubin Observatory. 
Furthermore, it would be beneficial to conduct real-time subthreshold analyses on neutrinos and x-ray transients. The Astrophysical Multimessenger Observatory Network (AMON) has attempted to achieve this~\cite{AyalaSolares:2019iiy,AyalaSolares:2023epo,AMON:2019zxe,AMONTeam:2020otr}. Enhancing the probability of discovering candidate sources of neutrino transients through searches for lower-energy neutrinos 
and/or weaker x-ray transients is a viable strategy.

\section{\label{sec:summary} Summary}

We presented diagnoses for the unified picture of high-energy neutrinos and UHECRs, focusing on $p\gamma$ scenarios. The framework is largely source independent and useful for determining the parameter space required for more sophisticated source models. Given that the UHECR energy generation rate density is comparable to that of high-energy neutrinos, the system must be semitransparent to the photomeson production, particularly if the spectrum is hard~\cite{Yoshida:2014uka}. This substantially limits the allowed model parameter space. Furthermore, the search for neutrino multiplets has excluded rare, bright source populations~\cite{Murase:2016gly}. As the production of neutrinos is inefficient if the luminosity of the source is too low, the cosmic-ray loading factor must be sufficiently high to satisfy the unification requirement for dim sources (see Fig.~\ref{fig:cr_loading_factor2D}). 
We investigated the impact of x-ray counterpart searches and demonstrated that the nondetection of x-ray emission allows us to place lower bounds on the cosmic-ray loading factor, which is a crucial quantity for source energetics (see Fig.~\ref{fig:cr_loading_factor2D_LL}). 
The ranges of magnetic fields and Lorentz factor are also limited depending on the source luminosity (see Fig.~\ref{fig:constraints_xiB_UHECR}).

We discussed different classes of astrophysical sources that can potentially satisfy the requirements for the unified models. Among them, LL GRBs and TDEs are
of particular interest as x-ray transients that can be tested for current and near-future x-ray missions. It is possible that realistic sources may not be standard candles. Therefore, it is necessary to obtain a sufficiently large sample to account for the diversity of source parameters in order to obtain reliable constraints. Nevertheless, our results demonstrate that real-time neutrino-triggered searches for x-ray transients are a powerful tool for testing the unified models for IceCube neutrinos and UHECRs. This is encouraging for current and future x-ray satellites with wide-field coverage, such as 
{\it MAXI}, {\it Einstein Probe}, {\it HiZ-GUNDAM}, and {\it Athena}. Neutrino detectors with bigger volumes or better angular resolutions are also beneficial for more efficient and frequent follow-up observation, for which KM3Net, Baikal-GVD, P-ONE, and especially IceCube-Gen2 will play important roles.

\begin{acknowledgments}
The authors are grateful to Wataru Iwakiri for his useful
input on the x-ray observations with {\it MAXI} and {\it NICER}. K.M. also thanks Bing Zhang for discussions during the 3rd Nanjing conference.  
This work by S.Y.is supported
by JSPS KAKENHI Grant No.~18H05206, 23H04892, and 
Institute for Advanced Academic Research
of Chiba University.
The work of K.M. is supported by the NSF Grants No.~AST-2108466, AST-2108467, and AST-2308021, and KAKENHI No.~20H01901 and 20H05852.
\end{acknowledgments}

\appendix
\onecolumngrid

\section{Analytical Formulas for calculating neutrino flux \label{apendix:analyticalFormulas}}
When we focus on the leading terms for simplicity in the integration of the formula in Eq.~(\ref{eq:general_yield_wz_sync}),
the neutrino luminosity 
originated from the target photon field below (above) the break energy is obtained by~\cite{Yoshida:2014uka}, 
\begin{equation}
\frac{d\dot{N}^{\rm L/H}_{\nu_e+\nu_\mu+\nu_\tau}}{d\varepsilon_{\nu}}\approx  
(\alpha_{\rm CR}-2)\xi_{\rm CR}\left(\frac{\varepsilon_p^{\rm FID}}{\varepsilon_{p0}[\Gamma]}\right)^{\alpha_{\rm CR}}
\frac{L_X^{\rm REF}}{(\varepsilon_p^{\rm FID})^2} 
\tau_0^{\rm L/H} \frac{3}{1-r_\pi}\frac{1}{x_\Delta^+-x_\Delta^-}f_{\rm sup}\mathcal{I}_{\rm P}^{\rm L/H}.
\label{eq:neutrino_yield_approx}
\end{equation}
Here $x_\Delta^{+(-)}$ is the maximal (minimal) bound of the relative energy of emitted pion
normalized by the parent cosmic-ray energy. A kinematic relation represents them
~\cite{Yoshida:2012gf} as
\begin{equation}
x_{\Delta}^{\pm} = \frac{(s_\Delta + m_{\pi}^2 - m_p^2) \pm \sqrt{(s_\Delta + m_{\pi}^2 - m_p^2)^2 - 4 s_\Delta m_{\pi}^2}}{2 s_\Delta}. 
\end{equation}
The neutrino energy term $\mathcal{I}_{\rm P}^{\rm L/H}$ is a power-law form in
most of the relevant neutrino energy ranges. We get

\begin{eqnarray}
\mathcal{I}_{\rm P}^{\rm L/H} = \left\{ 
\begin{array}{ll}
\frac{1}{(\alpha_{\rm CR}+2)^2} \left(\frac{\varepsilon_{\nu 0}^+[\varepsilon^{'\rm{L/H}}_{\rm min}]}{\varepsilon_{p0}[\Gamma]x_\Delta^+(1-r_\pi)}\right)^{(\alpha_X^{\rm L/H}+1)} \left(\frac{\varepsilon_\nu}{\varepsilon_{\nu 0}^+[\varepsilon^{'\rm{L/H}}_{\rm min}]}\right)^{-(\alpha_{\rm CR}+2)}, & ( \varepsilon_{\nu 0}^+[\varepsilon^{'\rm{L/H}}_{\rm min}] \leq  \varepsilon_\nu ) \nonumber  \\
\frac{1}{(\alpha_{\rm CR}+1-\alpha_X^{\rm L/H})^2} \left[\left(\frac{\varepsilon_\nu}{\varepsilon_{p0}[\Gamma]x_\Delta^+(1-r_\pi)}\right)^{-(\alpha_{\rm CR}+1-\alpha_X^{\rm L/H})} -\left(\frac{\varepsilon_{\nu 0}^+[\varepsilon^{'\rm{L/H}}_{\rm min}]}{\varepsilon_{p0}[\Gamma]x_\Delta^+(1-r_\pi)}\right)^{-(\alpha_{\rm CR}+1-\alpha_X^{\rm L/H})}\right]& \nonumber \\
+ \frac{1}{(\alpha_{\rm CR}+2)^2} \left(\frac{\varepsilon_{\nu 0}^+[\varepsilon^{'\rm{L/H}}_{\rm min}]}{\varepsilon_{p0}[\Gamma]x_\Delta^+(1-r_\pi)}\right)^{-(\alpha_{\rm CR}+1-\alpha_X^{\rm L/H})}, &( \varepsilon_{\nu 0}^+[\varepsilon^{'\rm{L/H}}_{\rm max}] \leq  \varepsilon_\nu \leq \varepsilon_{\nu 0}^+[\varepsilon^{'\rm{L/H}}_{\rm min}]) \nonumber  \\
    \frac{1}{(\alpha_{\rm CR}+1-\alpha_X^{\rm L/H})^2} \left[\left(\frac{\varepsilon_{\nu 0}^+[\varepsilon^{'\rm{L/H}}_{\rm max}]}{\varepsilon_{p0}[\Gamma]x_\Delta^+(1-r_\pi)}\right)^{-(\alpha_{\rm CR}+1-\alpha_X^{\rm L/H})} - \left(\frac{\varepsilon_\nu}{\varepsilon_{p0}[\Gamma]x_\Delta^-(1-r_\pi)}\right)^{-(\alpha_{\rm CR}+1-\alpha_X^{\rm L/H})}\right] & \nonumber \\
    + \frac{1}{(\alpha_{\rm CR}+1-\alpha_X^{\rm L/H})}\left(\frac{\varepsilon_{\nu 0}^+[\varepsilon^{'\rm{L/H}}_{\rm max}]}{\varepsilon_{p0}[\Gamma]x_\Delta^+(1-r_\pi)}\right)^{-(\alpha_{\rm CR}+1-\alpha_X^{\rm L/H})}\ln{\left(\frac{\varepsilon_{\nu 0}^+[\varepsilon^{'\rm{L/H}}_{\rm max}]}{\varepsilon_\nu}\right)},
    & ( \varepsilon_{\nu 0}^-[\varepsilon^{'\rm{L/H}}_{\rm max}] \leq  \varepsilon_\nu \leq \varepsilon_{\nu 0}^+[\varepsilon^{'\rm{L/H}}_{\rm max}]) \nonumber  \\
    \frac{1}{(\alpha_{\rm CR}+1-\alpha_X^{\rm L/H})}\left(\frac{\varepsilon_{\nu 0}^+[\varepsilon^{'\rm{L/H}}_{\rm max}]}{\varepsilon_{p0}[\Gamma]x_\Delta^+(1-r_\pi)}\right)^{-(\alpha_{\rm CR}+1-\alpha_X^{\rm L/H})}\ln{\left(\frac{x_\Delta^+}{x_\Delta^-}\right)},
    & (\varepsilon_\nu \leq  \varepsilon_{\nu 0}^-[\varepsilon^{'\rm{L/H}}_{\rm max}]). \\
\end{array}
\right. .
\label{eq:neutrino_energy_dist}
\end{eqnarray}
Here, $\varepsilon_{\nu 0}^\pm[\varepsilon'_X]$
is given by
\begin{equation}
\varepsilon_{\nu 0}^\pm[\varepsilon'_X] = \frac{s_\Delta - m_p^2}{4\varepsilon'_X}\Gamma x_\Delta^\pm(1-r_\pi).
\label{eq:neutrino_energy_bound}
\end{equation}
This parametrization is understood by the fact that
the main energy range of neutrinos produced by the photon energy $\varepsilon'_X$
is approximately represented by $\varepsilon_{\nu 0}^-[\varepsilon'_X] \leq \varepsilon_\nu \leq \varepsilon_{\nu 0}^+[\varepsilon'_X]$.

$\varepsilon_{\rm min/max}^{'\rm L/H}$ specified the photon energy range (in the plasma rest frame) in the BPL spectrum
represented by Eq.~(\ref{eq:target_photon}). They are given by
\begin{eqnarray}
    \varepsilon^{'\rm L}_{\rm min}  =  \varepsilon'_{\rm min} &,&\quad \varepsilon^{'\rm L}_{\rm max}  =  \varepsilon'_{X,b} \nonumber\\
    \varepsilon^{'\rm H}_{\rm min}  =  \varepsilon'_{X,b} &,&\quad \varepsilon^{'\rm H}_{\rm max}  =  \varepsilon'_{\rm max} 
\end{eqnarray}

The total differential neutrino yield emitted from a source is then given by
\begin{equation}
\frac{d\dot{N}_{\nu_e+\nu_\mu+\nu_\tau}}{d\varepsilon_{\nu}} = \frac{d\dot{N}^{\rm L}_{\nu_e+\nu_\mu+\nu_\tau}}{d\varepsilon_{\nu}} + \frac{d\dot{N}^{\rm H}_{\nu_e+\nu_\mu+\nu_\tau}}{d\varepsilon_{\nu}}
\end{equation}

Most of the parameter space relevant to the present sensitivities of x-ray observation facilities
involves optically thin sources. UHECR emitters also must be optically thin. Nevertheless,
we introduced the calorimetric limit in the calculations of the neutrino flux by employing the limit
on the optical depth as
\begin{equation}
\tau_0^{\rm L/H}\leq 1/\kappa_{p\gamma} \approx 5
\label{eq:calorimetric}
\end{equation}
where $\kappa_{p\gamma}\sim0.2$ is the proton inelasticity. For the parameter set beyond this limit,
we placed $\tau_0=5$ to calculate the neutrino yield as the calorimetric condition.

When the target photon spectrum follows the thermal spectrum,
the neutrino yield is obtained by following Eq.~(7) in Ref~\cite{Yoshida:2012gf} 
with minor modifications and reparametrization, and given by
\begin{equation}
    \frac{d\dot{N}_{\nu_e+\nu_\mu+\nu_\tau}}{d\varepsilon_{\nu}}\approx 
(\alpha_{\rm CR}-2)\xi_{\rm CR}\left(\frac{\varepsilon_p^{\rm FID}}{\varepsilon_{p0}[\Gamma]}\right)^{\alpha_{\rm CR}}
\frac{L_X^{\rm REF}}{(\varepsilon_p^{\rm FID})^2}
\frac{3}{1-r_\pi}\frac{1}{x_\Delta^+-x_\Delta^-}f_{\rm sup}\tau_{p\gamma}(\varepsilon_\nu).
\label{eq:neutrino_yield_thermal}
\end{equation}
where the optical depth $\tau_{p\gamma}$ is given by
\begin{equation}
    \tau_{p\gamma}(\varepsilon_\nu)=
    \frac{15}{32\pi^5}\sqrt{\frac{L_X}{\xi_{\rm B}}}\frac{B'}{\Gamma^2}
    \frac{1}{(k_{\rm B}T)^3\sqrt{2c}}
    \int\limits^{\infty}_{\frac{\varepsilon_\nu}{(1-r_\pi)x^+_\Delta}} 
    d\varepsilon_p\frac{\Gamma^3}{\varepsilon_p^3}
    \left(\frac{\varepsilon_p}{\varepsilon_{p0}[\Gamma]}\right)^{-\alpha_{\rm CR}} F(s,\varepsilon_p)
\end{equation}
\begin{eqnarray}
    F(s,\varepsilon_p) &\equiv& \int ds (s-m_p^2) \sigma_{p\gamma}(s)\ln{\left(\frac{x_\Delta^+}{\xi_p}\right)}\{-\ln{(1-e^{-\Gamma\frac{s-m_p^2}{4k_{\rm B}T\varepsilon_p}})}\} \label{eq:s_factor}\\
    &\approx& (s_\Delta-m_p^2)\Delta_{s_\Delta} \sigma_{p\gamma}^\Delta\ln{\left(\frac{x_\Delta^+}{\xi_p}\right)}\{-\ln{(1-e^{-\Gamma\frac{s_\Delta-m_p^2}{4k_{\rm B}T\varepsilon_p}})}\} \label{eq:s_factor_approx}
\end{eqnarray}
\begin{equation}
\xi_p = \begin{cases}
x_\Delta^-                         & \varepsilon_\nu \leq (1-r_\pi)x_\Delta^{-} \varepsilon_p,\\
{\varepsilon_\nu\over (1-r_\pi)\varepsilon_p} & \text{otherwise}.
\end{cases}
\label{eq:xi_range}
\end{equation}
The simplification from Eq.~(\ref{eq:s_factor}) to Eq.~(\ref{eq:s_factor_approx}) is based on the $\Delta$ resonance approximation.

\section{UHECR-Neutrino common Sources with thermal x-ray emission}

The x-rays in the cosmic-ray acceleration/neutrino production region 
can be thermalized if the environment is optically thick at the initial phase
of the system. A choked-jet scenario~\cite{Murase:2013ffa, Senno:2015tsn} 
is a good example of this possibility.
As long as we are considering the UHECR-neutrino common sources, 
the emission region must still be optically thin when emitting UHECRs and neutrinos. 
Here, we briefly review our modeling of the optically thin thermal x-ray source in a generic way 
to see if the outcomes we presented in the earlier sections are solid or not.

\begin{figure}
  \begin{center}
  \includegraphics[width=0.5\textwidth]{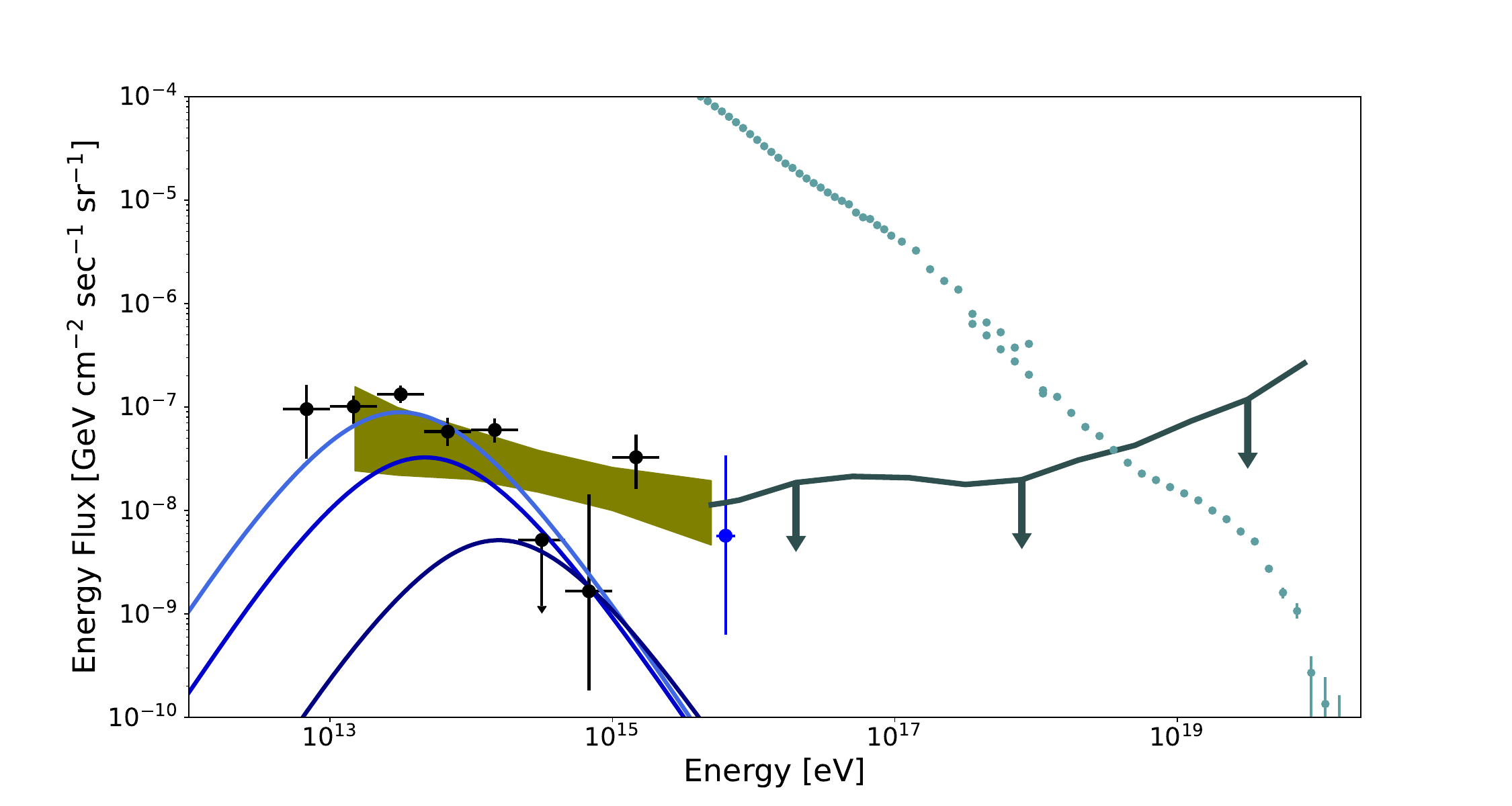}
  \caption{Same as Fig.~\ref{fig:neutrino_diffuse_fluxes}, but the thermal photon case.
  The three cases of $\Gamma$ values ,$\Gamma=2$ (top), $\Gamma=3$ (middle), $\Gamma=10$ (bottom)
  are presented.}
  \label{fig:neutrino_diffuse_fluxes_thermal}
  \end{center}
\end{figure}

\begin{figure*}
  \begin{center}
  \includegraphics[width=0.48\textwidth]{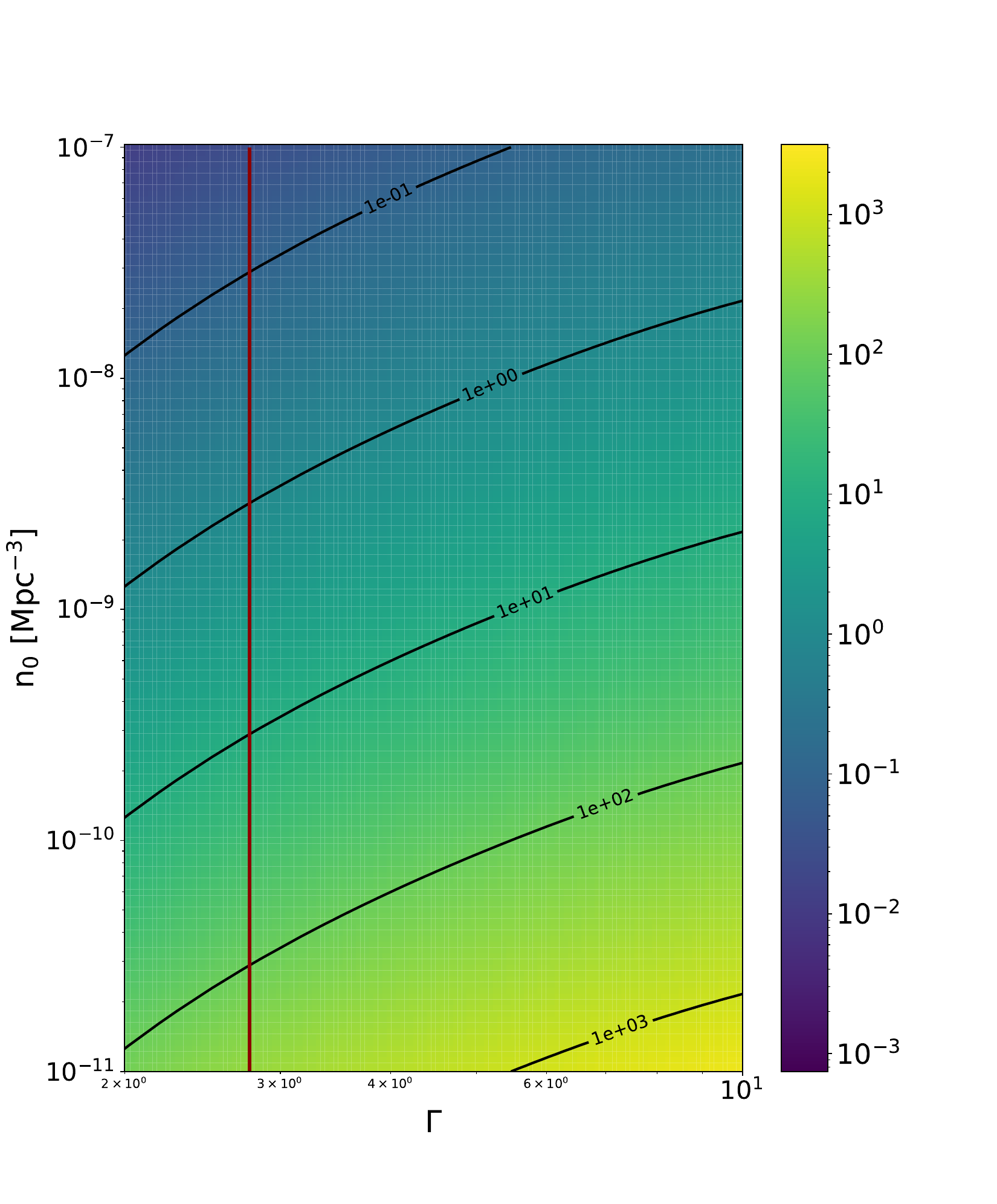}
  \includegraphics[width=0.48\textwidth]{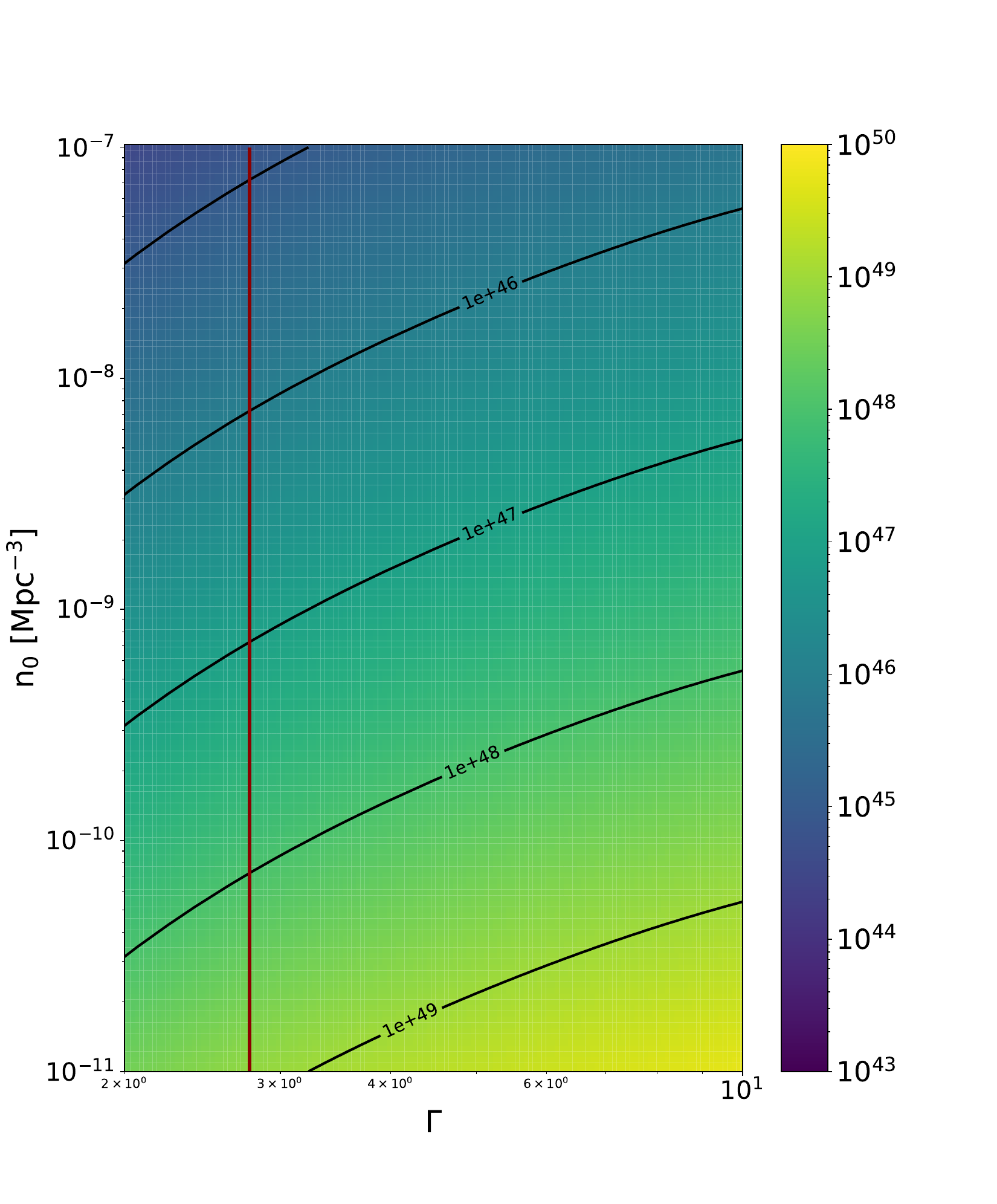}\\
  \caption{Same as Fig.~\ref{fig:cr_loading_factor2D}, but the case when the target photon spectrum
  follows the thermal distribution given by Eq.~(\ref{eq:thremal_photon_spectrum}).
  The x-ray thermal energy is $k_{\rm B}T=0.177 {\rm keV}$ for corresponding to 
  $\varepsilon'_{X,b}=0.5 {\rm keV}$ in the nonthermal emission case.}
  \label{fig:cr_loading_factor2D_thermal}
  \end{center}
\end{figure*}

The target x-ray photons are now assumed to follow the quasi-Planck distribution 
instead of the BPL written as Eq.~(\ref{eq:target_photon}). It is given by
\begin{equation}
    \frac{dn_X}{d\varepsilon_X'} = \frac{\epsilon_{\rm T}}{\pi^2\hbar^3 c^3}
    \frac{\varepsilon^{'2}_X}{\exp^{\frac{\varepsilon'_X}{k_{\rm B}T}}-1}
    \label{eq:thremal_photon_spectrum}
\end{equation}
$\epsilon_{\rm T}=1$ corresponds to the blackbody distribution

The x-ray luminosity density is associated with the thermal photon energy density,
\begin{equation}
    \frac{L'_X}{4\pi R^2 c} = \epsilon_{\rm T} \frac{\pi^2}{15\hbar^3 c^3}(k_{\rm B}T)^4
\end{equation}
$R$ is the distance of an acceleration and emission region from the central engine.

The x-ray density is also associated with the magnetic energy density via $\xi_{\rm B}$
and we get
\begin{equation}
    \frac{B^{'2}}{8\pi} = \xi_{\rm B}\frac{L'_X}{4\pi R^2 c}
\end{equation}

We then get the relations between $\epsilon_{\rm T}$, $B'$, and $\xi_{\rm B}$
as
\begin{eqnarray}
\varepsilon_{\rm T} &=& \frac{15\hbar^3c^3}{8\pi^3}\frac{B^{'2}}{(k_{\rm B}T)^4}\xi_{\rm B}^{-1}\nonumber\\
&\approx& 4.6\times 10^{-10}\left(\frac{k_{\rm B}T}{0.5{\rm keV}}\right)^{-4}
\left(\frac{B'}{100~{\rm G}}\right)^2\left(\frac{\xi_{\rm B}}{0.1}\right)^{-1}\nonumber\\
\label{eq:glay_body_scale}
\end{eqnarray}
It implies that the reasonable range of $\xi_B$ and $B'$ 
requires $\epsilon_{\rm T}\lesssim 10^{-9}\ll 1$,
providing the optically thin environment.

Hereafter we consider the case when $k_{\rm B}T = 0.355\varepsilon'_{X,b}=0.177$~keV
so that its peak energy is close to $\varepsilon'_{X,b}$.
The analytical formulas to calculate the neutrino yield are described in Appendix A
and given by Eq.~(\ref{eq:neutrino_yield_thermal}).
Figure~\ref{fig:neutrino_diffuse_fluxes_thermal} shows the resultant
neutrino background flux with the same baseline parameters as Fig.~\ref{fig:neutrino_diffuse_fluxes}.
The spectrum gets rapidly steeper at energies beyond the peak energy $\varepsilon_\nu^{\rm peak}$ 
which corresponds to the representative energy of neutrinos produced by protons with
energies of $\varepsilon_p\simeq \varepsilon_{p0}[\Gamma]$.
Although the introduced analytical $\Delta$-resonance approximation artificially enhances
this feature, it implies that this scenario contributes mostly to the cosmic background flux 
in the energy region below 100 TeV. We, thus, scan the range of cosmic-ray loading factor $\xi_{\rm CR}$
to meet 
$\displaystyle 2\times 10^{-8} \leq E_\nu^2\Phi_{\nu_e+\nu_\mu+\nu_\tau}|_{E_\nu=\varepsilon_\nu^{\rm peak}}
\leq 2\times 10^{-7}$ [GeV cm$^{-2}$ sec$^{-1}$ sr$^{-1}$]. The resultant upper bound of $\xi_{\rm CR}$
and the corresponding $L_{\rm UHECR}$ are displayed in Fig.~\ref{fig:cr_loading_factor2D_thermal}.
The constraints are similar to the nonthermal BPL case summarized in Fig.~\ref{fig:cr_loading_factor2D}.
The conclusion remains the same, given that the thermal x-ray field 
in the neutrino production zone is optically thin,
determined by $B'$ and $\xi_{\rm B}$ as represented in Eq.~(\ref{eq:glay_body_scale}).

\twocolumngrid

\bibliography{syoshida}

\end{document}